\documentclass[10pt,letterpaper]{article}
\usepackage[T1]{fontenc}
\usepackage[utf8]{inputenc}
\usepackage[letterpaper,left=0.875in,right=0.875in,top=0.9in,bottom=1.1in,columnsep=0.25in]{geometry}
\usepackage{times,microtype,graphicx,booktabs,array,longtable,ragged2e}
\usepackage[table]{xcolor}
\usepackage{amsmath,amssymb,textcomp,newtxmath}
\usepackage[numbers,sort&compress]{natbib}
\usepackage{caption,fancyhdr,titlesec}
\usepackage{pdflscape,placeins,enumitem,csvsimple-l3,xurl}
\usepackage[normalem]{ulem}
\usepackage[hidelinks,unicode,hyperfootnotes=false]{hyperref}
\newcommand{\PaperTitle}{V2G: An Agentic Pipeline for Gene Prioritisation}
\newcommand{\RunningTitle}{V2G: An Agentic Pipeline for Gene Prioritisation}
\newcommand{\EmailOne}{\href{mailto:shakuntala.baichoo@uhn.ca}{\nolinkurl{shakuntala.baichoo}}}
\newcommand{\EmailTwo}{\href{mailto:sohail.sarkar@uhn.ca}{\nolinkurl{sohail.sarkar}}}
\newcommand{\EmailThree}{\href{mailto:bowang@vectorinstitute.ai}{\nolinkurl{bowang@vectorinstitute.ai}}}
\makeatletter
\newcommand{\AuthorAffiliations}{%
  \begingroup
  \let\thefootnote\relax
  \renewcommand{\@makefntext}[1]{\noindent##1}%
  \footnotetext{%
    \textsuperscript{1}PMCC AI Lab, University Health Network;\\
    \textsuperscript{2}Vector Institute\\
    Emails: \{\EmailOne,\EmailTwo\}@uhn.ca;\\
    \EmailThree.}%
  \endgroup}
\makeatother
\hypersetup{pdftitle={\PaperTitle},
  pdfauthor={Shakuntala Baichoo, Sohail(Neel) Sarkar, Bo Wang},
  pdfsubject={Preprint},
  pdfkeywords={GWAS, gene prioritisation, language models, evidence, recall}}
\AddToHook{env/landscape/begin}{%
  \clearpage
  \pagestyle{empty}%
  \AddToHook{shipout/foreground}[landscape-page-number]{%
    \put(\dimexpr\paperwidth-0.5in\relax,-0.5\paperheight){%
      \rotatebox{90}{\makebox(0,0){\normalfont\normalsize\thepage}}}}}
\AddToHook{env/landscape/end}{%
  \clearpage
  \RemoveFromHook{shipout/foreground}[landscape-page-number]}
\definecolor{tablehead}{gray}{0.94}
\definecolor{tablehighlight}{gray}{0.97}
\setcitestyle{numbers,square,comma,sort&compress}

\setlist{nosep,leftmargin=1.6em}
\titleformat{\section}{\large\bfseries}{\thesection.}{0.5em}{}
\titleformat{\subsection}{\normalsize\bfseries}{\thesubsection.}{0.5em}{}
\titleformat{\subsubsection}{\normalsize\bfseries}{\thesubsubsection.}{0.5em}{}
\titlespacing*{\section}{0pt}{10pt}{3pt}
\titlespacing*{\subsection}{0pt}{8pt}{2pt}
\titlespacing*{\subsubsection}{0pt}{6pt}{2pt}
\renewenvironment{abstract}{%
  \begin{center}\bfseries Abstract\end{center}%
  \begin{list}{}{\setlength{\leftmargin}{0.15in}\setlength{\rightmargin}{0.15in}}%
  \item\relax\small}{\end{list}\vspace{0.5em}}
\begin{document}
\twocolumn[{
\begin{center}
\rule{\textwidth}{0.8pt}\\[0.6em]
{\LARGE\bfseries\PaperTitle\par}
\vspace{0.6em}
\rule{\textwidth}{0.5pt}\\[0.7em]
{\normalsize\bfseries Shakuntala Baichoo\textsuperscript{1}\qquad
Sohail (Neel) Sarkar\textsuperscript{1}\qquad Bo Wang\textsuperscript{2}\par}
\end{center}
\vspace{0.6em}
}]
\AuthorAffiliations
\thispagestyle{plain}

\begin{abstract}
Assigning GWAS variants to causal genes requires evidence from multiple resources. We present V2G, an agentic pipeline that records evidence retrieval and returns ranked genes alongside the retrieved evidence. We evaluate whether language-model reranking improves gene prioritisation and whether its gains depend on the retrieved evidence or recall of known associations. On 135 held-out loci where candidate retrieval succeeded and Open Targets L2G predicts, reranking by a language model improves Precision@1 over the evidence ordering it receives, 0.793 against 0.711 (+0.081, 95\% CI +0.038 to +0.129), and over nearest-gene assignment (+0.178); no difference from Open Targets L2G or from training-membership baselines was detected. Given only the trait and the candidate gene symbols, in shuffled order, the same model scores 0.864 on these loci (three seeds, 0.859--0.867) and names the reference gene at 0.594 of the loci where the pipeline never retrieved it; with each gene’s symbol moved to another gene’s evidence row it follows the top-scored row and scores 0.284. The gain over the evidence ordering is recall of known genes, and the evidence as presented anchors the model to the ordering rather than informing it. Across the 296 held-out loci (five seeds each) the model keeps the given top candidate at 86\% of decisions; at naturally retrieved loci its departures select a reference gene in 85\% of departure decisions (31 of 37 loci), and none displaced one. Its calls carry trait-matched mouse mutant or Mendelian phenotype annotations more often than other candidates at the same locus. Train--test overlap is reported throughout. The study includes a frozen benchmark, cached predictions and reproduction code.
\end{abstract}

\section{Introduction}

Genome-wide association studies have mapped tens of thousands of trait-associated loci; however, most associated variants are non-coding and act through regulation rather than protein alteration~\citep{maurano2012,canogamez2020}. Deciding which gene at a locus carries the signal stands between an association and a mechanistic or therapeutic hypothesis, and it matters downstream: drug targets with genetic support are more likely to succeed in development~\citep{nelson2015}. Proximity can mislead; at the \emph{FTO} obesity locus, the associated variants act on the more distant \emph{IRX3}~\citep{smemo2014}, and linkage disequilibrium routinely leaves several plausible candidates within a few hundred kilobases~\citep{schaid2018}.

The evidence needed to resolve a locus is not scarce but scattered. Expression QTLs~\citep{gtex2020}, colocalization posteriors~\citep{giambartolomei2014}, chromatin contacts~\citep{fulco2019}, coding consequences~\citep{mclaren2016}, Mendelian randomization estimates~\citep{daveysmith2003}, and population allele frequencies~\citep{karczewski2020} each live in a different resource, with different coordinates, different coverage, and different failure modes. Expression QTLs, for instance, are systematically least informative for precisely the genes that GWAS implicates~\citep{mostafavi2023}. Existing methods address this by compressing the available evidence into a single number. Gene-level association methods aggregate variant statistics within gene boundaries~\citep{deleeuw2015} or impute expression and test it against the trait~\citep{gamazon2015,gusev2016,barbeira2018}. Integrative scores combine regulatory and molecular evidence into one ranking: the Open Targets locus-to-gene score trains a supervised model on curated loci~\citep{mountjoy2021,ochoa2023}, combined-SNP-to-gene aggregates orthogonal functional links~\citep{gazal2022}, and polygenic priority scores exploit genome-wide enrichment of gene features~\citep{weeks2023}.

A single score is convenient and, in our experience of using these tools, not always sufficient. Integrative scores increasingly expose their inputs (Open Targets now reports the feature values behind each L2G prediction and their contributions to it~\citep{buniello2025}) but what is explained is a fixed model’s features, not an argument for one gene over another that a user can check against the retrieved evidence locus by locus. Nor does a score indicate where the method declined to answer. Both matter in practice because the output of variant-to-gene assignment is not an end in itself but the input to an experiment someone has to fund.

We take a different approach to the same evidence. Rather than compressing it, an agent retrieves each line as a separate, recorded tool call (fine-mapping the locus, querying expression QTLs, testing colocalization, computing Mendelian randomization, and reading per-population frequencies) and then reasons over the assembled evidence in language, citing what it used. Language-model agents coupled with domain tools have been applied to biomedical retrieval tasks of this shape~\citep{jin2024genegpt,gao2024}. The output is a ranked gene with a rationale that can be inspected, accepted, or rejected on its merits. Whether that structure buys accuracy, as opposed to transparency alone, is an empirical question, and one that is harder to answer than it appears.

This is more difficult because the benchmarks for this task are fragile. Curated gene--trait gold standards are assembled from the literature, which also informs the methods being tested, so recall can masquerade as inference. A benchmark must decide what to do when its evidence pipeline fails to retrieve the gold gene, with the usual choice of adding it to the candidate set and labelling that gene by its own absence. Both failure modes are visible in the results reported below: two lookups that rank by membership in the training reference set are not distinguishable from any method we test on the primary analysis, and a classifier given only the pattern of missing features is perfect on exactly the loci whose reference gene was added, and both were first reported in a companion manuscript~\citep{companion2026}. We report every result under them: force-added and naturally retrieved loci separately throughout, and both a coverage-conditional and a coverage-adjusted view of any method that may decline to answer the question.

We evaluate whether language-model reranking improves gene prioritisation beyond the evidence ordering and existing methods, and whether its departures from that ordering select reference genes. Prompt ablations test the model's dependence on gene names and the presented evidence. Within-locus phenotype comparisons assess functional support for the selected genes.

\section{Results}\label{sec:results}

\subsection{An agentic pipeline that returns evidence, not only a score}\label{sec:an-agentic-pipeline-that-returns-evidence-not-only-a-score}

\begin{figure*}[!t]
\centering
\includegraphics[width=1\textwidth,height=0.49\textheight,keepaspectratio]{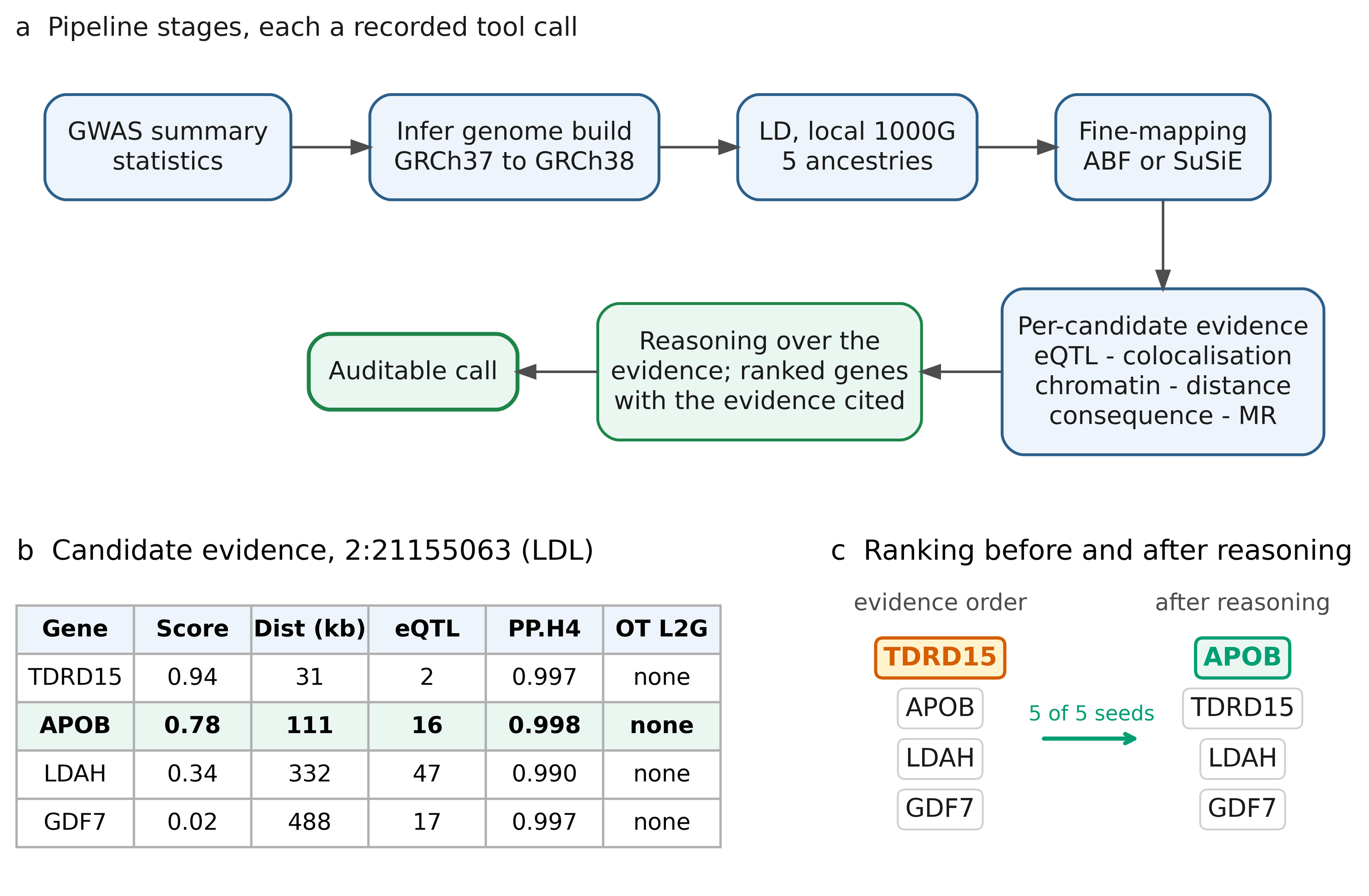}
\caption{\textbf{An agentic variant-to-gene pipeline and a locus where the evidence ordering is incorrect.} \textbf{a}, Pipeline stages. Each is a separate, recorded tool call: the genome build is inferred from the coordinates and GRCh37 input lifted to GRCh38; linkage disequilibrium is computed from local 1000 Genomes panels across five ancestry groups; loci are fine-mapped to credible sets by approximate Bayes factor or SuSiE; per-candidate evidence is retrieved gene by gene; and a language model reasons over the assembled evidence and returns a ranked list citing what it used, so any call can be traced to the measurements behind it. \textbf{b}, The evidence retrieved at 2:21155063 A>G, associated with LDL cholesterol (study GCST90002412, a held-out test locus). Score is the composite evidence score that determines the order in which candidates are presented; Dist is the distance from the lead variant to the gene’s transcription start site; eQTL is the number of GTEx v8 tissues with a significant eQTL; PP.H4 is the posterior probability of colocalization; OT L2G is the Open Targets locus-to-gene score. The curated gold gene, APOB, is shaded. All four candidates colocalize at PP.H4~\ensuremath{\geq}~0.99; therefore, colocalization cannot discriminate between them, and Open Targets returns no L2G prediction at this locus. \textbf{c}, Ranking before and after the reasoning step. The composite score ranks TDRD15 first on proximity (31~kb, against 111~kb for APOB); the model promotes APOB in all five seeds, with rationales that cite its role in LDL particles (prior knowledge rather than a retrieved measurement). With the open-weights backbone (Qwen3-32B) the ordering is left unchanged in all five seeds. The values shown are the stored benchmark evidence, reproduced in \hyperref[tab:s7]{Supplementary Table~7}. Of these, the evaluated prompt showed the score and the eQTL and colocalisation flags; the distance, tissue count, PP.H4 and L2G columns were not shown (Methods).}\label{fig:1}
\end{figure*}

The pipeline takes GWAS summary statistics and returns ranked candidate genes with the evidence behind each call (Fig.~\ref{fig:1}a). It first infers the genome build from the coordinates themselves and converts GRCh37 input to GRCh38~\citep{kent2002,hapmap2010}, so that every downstream resource agrees; computes linkage disequilibrium from local 1000 Genomes reference panels across five ancestry groups and two builds~\citep{tgp2015}; and fine-maps each locus to credible sets by approximate Bayes factor~\citep{wakefield2009} or SuSiE~\citep{wang2020susie}. For each candidate gene, it then issues discrete, recorded queries for expression QTLs~\citep{gtex2020}, colocalization posteriors~\citep{giambartolomei2014}, chromatin and distance evidence, variant consequences~\citep{mclaren2016}, per-population allele frequencies11, and two-sample Mendelian randomization estimates~\citep{burgess2013}. A language model then reasons over the assembled evidence for the locus and returns a ranked list with the evidence it relied on, so that any call can be traced back to the measurements that produced it.

A single locus shows what this buys. At 2:21155063 A>G, associated with LDL cholesterol (Fig.~\ref{fig:1}b,~c), four genes are retrieved. The composite evidence score ranks \emph{TDRD15} first, at 31~kb from the lead variant, ahead of \emph{APOB} at 111~kb. All four candidates colocalise at posterior probability \ensuremath{\approx}~0.99, so colocalization cannot discriminate between them, and Open Targets returns no L2G prediction at this locus at all. The reasoning step promotes \emph{APOB}, the curated reference gene, in all five seeds. Its rationales rest on prior knowledge of \emph{APOB} as the structural protein of LDL particles rather than on a retrieved measurement: the model was shown an eQTL flag, not tissues, and the stored top-ten eQTL tissue list for \emph{APOB} does not include the liver. At this locus, the evaluated prompt showed, per gene, the composite score and the eQTL and colocalization flags, which were set alike for all four candidates; no distance, tissue, posterior, or L2G value was shown (Methods; the prompt is reproduced in \hyperref[note:7]{Supplementary Note~7}). The example shows a correction that draws on biological knowledge the aggregate score lacks; because the \emph{APOB}--LDL association is familiar, it does not show inference beyond recall. The uniform colocalization posteriors are discussed in the Methods section.

\subsection{A benchmark that separates recall from inference}\label{sec:a-benchmark-that-separates-recall-from-inference}

We evaluated 844 fine-mapped loci spanning 180 traits and 175 EFO terms drawn from the GWAS Catalog~\citep{cerezo2025} and split 548 training/296 test loci. The split was intended to hold out gene--trait associations, and it does so only partly: 87 test loci share a reference gene and exact trait name with a training locus, 99 share a reference gene and EFO term, 154 share a reference gene with some training locus, and 7 share the lead variant, leaving 197 test loci with no gene--EFO overlap (\hyperref[tab:s8]{Supplementary Table~8}). A further 88 replicate loci, not counted in the 844, from independent cohorts (Finnish 35, East Asian 19, South Asian 17, African 8, Greater Middle Eastern 4, Hispanic or Latin American 3, Central Asian 1, European 1)~\citep{martin2019} all share a reference gene--EFO pair with training, so they test whether a known association transfers between cohorts and cannot separate transfer from recall; at 20 of them no reference gene is among the candidates, so top-1 accuracy there cannot exceed 68/88. Candidate sets contain a median of four genes.

Two properties of such benchmarks, first reported in a companion manuscript~\citep{companion2026} and reproduced in the results below, constrain everything that follows. Curated gold standards reward recall of associations already in the literature. And gold genes that the evidence pipeline fails to retrieve must be added to the candidate set for the locus to remain scorable, which marks them by the pattern of evidence they lack. We therefore report the 114 loci with a force-added reference gene separately from the 182 natural loci, and take as the primary analysis the 135 natural loci at which Open Targets also returns a prediction. That analysis is conditional: it measures ranking at 45.6\% of test loci where candidate retrieval succeeded and Open Targets has coverage, and it is not a common coverage set for the gene-level, transcriptome-wide, and SNP-to-gene comparators, which answer at 50, 42, and 111 of these loci. The corrected candidate sets are released.

\subsection{Accuracy: better than the evidence ordering; no difference from Open Targets L2G detected}\label{sec:accuracy-better-than-the-evidence-ordering-no-difference-from-open-targets-l2g-detected}

On the primary split, the pipeline reached a precision @1 of 0.793 (107/135; Wilson 95\% CI 0.717--0.852), against 0.778 for gene membership with evidence-score tie-breaking, 0.763 for Open Targets L2G~\citep{mountjoy2021}, 0.763 for trait membership with evidence-score tie-breaking, 0.719 for a gradient-boosted model~\citep{chen2016xgboost} trained on naturally retrieved loci, 0.711 for the evidence ordering the model is given, and 0.615 for nearest-gene assignment (Table~\ref{tab:1}). Paired on the same loci (exact McNemar test~\citep{mcnemar1947}, unadjusted; 95\% intervals from a bootstrap that keeps loci sharing a reference gene together), the pipeline improved on the evidence ordering at 11 loci and was worse at none (difference +0.081, +0.038 to +0.129; \emph{P}~=~9.8~\ensuremath{\times}~10\textsuperscript{\ensuremath{-}4}) and on nearest-gene assignment 35 to 11 (+0.178, +0.079 to +0.280; \emph{P}~=~5.4~\ensuremath{\times}~10\textsuperscript{\ensuremath{-}4}); against the gradient-boosted model, it was 16 to 6 (+0.074, +0.007 to +0.143; \emph{P}~=~0.052). No difference was detected against Open Targets L2G (12 to 8; +0.030, \ensuremath{-}0.038 to +0.098; \emph{P}~=~0.50), gene membership (12 to 10; P~=~0.83), or trait membership (11 to 7; \emph{P}~=~0.48) (Fig.~\ref{fig:2}c). Not detecting a difference is not equivalence: these intervals admit differences of several percentage points in either direction, and no equivalence margin was set.

\begin{table*}[!t]
\centering
\caption{\textbf{Precision@1 by method and split.} Force-added loci (reference gene added to the candidate set by the benchmark and so buried by construction) are reported separately from natural loci, with the pooled result alongside. The primary analysis is the natural loci where Open Targets also returns a prediction; it is conditional on candidate retrieval and Open Targets coverage. The lower two rows are diagnostics, not comparators: both reach 1.000 on force-added loci, which is the artifact those loci create rather than a result. Open Targets is scored with uncovered loci as failures; the other estimands are in Table~\ref{tab:2}. Every rate is rounded once from its count; the full grid with counts and Wilson intervals is \hyperref[tab:s1]{Supplementary Table~1}.}\label{tab:1}
\fontsize{8.5}{10}\selectfont
\setlength{\tabcolsep}{3pt}
\renewcommand{\arraystretch}{1.12}
\begin{tabular}{@{}>{\RaggedRight\arraybackslash}p{\dimexpr0.39000\linewidth-2\tabcolsep\relax}>{\RaggedRight\arraybackslash}p{\dimexpr0.09000\linewidth-2\tabcolsep\relax}>{\RaggedRight\arraybackslash}p{\dimexpr0.10000\linewidth-2\tabcolsep\relax}>{\RaggedRight\arraybackslash}p{\dimexpr0.11000\linewidth-2\tabcolsep\relax}>{\RaggedRight\arraybackslash}p{\dimexpr0.11000\linewidth-2\tabcolsep\relax}>{\RaggedRight\arraybackslash}p{\dimexpr0.10000\linewidth-2\tabcolsep\relax}>{\RaggedRight\arraybackslash}p{\dimexpr0.10000\linewidth-2\tabcolsep\relax}@{}}
\toprule
\rowcolor{tablehead}
\textbf{Method} & \textbf{All} & \textbf{Natural} & \textbf{Force-added} & \textbf{OT predicts} & \textbf{OT scores all} & \textbf{Primary} \\
\rowcolor{tablehead}
 & (296) & (182) & (114) & (219) & (102) & \textbf{(135)} \\
\midrule
\begin{filecontents*}{tables/table_1.tsv}
Shade	CellA	CellB	CellC	CellD	CellE	CellF	CellG
	Agentic pipeline (GPT-5.2, consensus)	0.486	0.753	0.061	0.516	0.725	\textbf{0.793}
	Gene membership + evidence tie-break	0.611	0.769	0.360	0.603	0.735	0.778
	Open Targets L2G	0.365	0.566	0.044	0.493	0.735	0.763
	Trait membership + evidence tie-break	0.574	0.742	0.307	0.575	0.725	0.763
	GBM trained on naturally retrieved loci	0.409	0.632	0.053	0.470	0.657	0.719
	Evidence ordering (as given to the model)	0.395	0.621	0.035	0.457	0.667	0.711
	Agentic pipeline (Qwen3-32B, consensus)	0.395	0.621	0.035	0.457	0.667	0.711
	Gene frequency (no evidence)	0.659	0.665	0.649	0.616	0.627	0.637
	Nearest gene (GENCODE TSS)	0.372	0.549	0.088	0.402	0.598	0.615
	Gene--trait frequency (no evidence)	0.530	0.566	0.474	0.530	0.569	0.548
	\emph{Diagnostics (not comparators)}						
	GBM trained on all loci	0.787	0.654	1.000	0.849	0.765	0.756
	GBM, missingness only (control)	0.740	0.577	1.000	0.781	0.647	0.644
\end{filecontents*}
\csvreader[separator=tab,head to column names,late after line=\\]{tables/table_1.tsv}{}{\Shade\CellA & \Shade\CellB & \Shade\CellC & \Shade\CellD & \Shade\CellE & \Shade\CellF & \Shade\CellG}
\bottomrule
\end{tabular}
\end{table*}

A classifier given exactly the inputs the model saw (the composite score and the three flags) and trained on the naturally retrieved training loci does not reproduce the gain: on the primary split a gradient-boosted model scores 0.644 and a logistic regression 0.689, both below the evidence ordering itself (0.711), and the pipeline is better than each (25 loci to 5, \emph{P}~=~3.2~\ensuremath{\times}~10\textsuperscript{\ensuremath{-}4}; 14 to 0, \emph{P}~=~1.2~\ensuremath{\times}~10\textsuperscript{\ensuremath{-}4}; \hyperref[tab:s10]{Supplementary Table~10}). Whatever the model adds at these loci was not recovered by either classifier from the four fields it was shown.

Withholding the evidence raises accuracy. Given the trait and the candidate symbols alone, in shuffled order, GPT-5.2 selects the reference gene at 0.864 of the primary loci (mean of three seeds; 0.859--0.867; 116--117/135; Wilson 95\% CI 0.79--0.91 at each seed), above the full prompt of the same seed at every seed (0.778--0.807; 7 loci to 15--18, \emph{P}~=~0.134, 0.093 and 0.043), the five-seed consensus (0.793; 7 to 16--17, \emph{P}~=~0.064--0.093) and the evidence ordering (0.711; 7 to 27--28, \emph{P}~\(\leq\)~8.3~\ensuremath{\times}~10\textsuperscript{\ensuremath{-}4}). Adding the three binary flags changes nothing on the primary split (0.862; 0.852--0.867; 2--3 loci to 2, \emph{P}~=~1.0 at each seed). On the 97 primary loci whose reference gene and trait never occur together among the training reference genes, the names-only prompt scores 0.832 (0.825--0.835) against 0.794 (0.783--0.804) for the full prompt (6 loci to 9--10, \emph{P}~=~0.45--0.61) and 0.701 for the evidence ordering (6 to 18--19, \emph{P}~\(\leq\)~0.023). On the 182 naturally retrieved loci, the names-only prompt scores 0.839 (0.835--0.846) against 0.758 (0.742--0.769) for the full prompt (9--11 loci to 23--26, \emph{P}~=~0.058, 0.014 and 0.006) and 0.753 for the consensus (\emph{P}~=~0.006--0.020). On the 114 force-added loci, where the reference gene was never retrieved and entered the candidate set with a placeholder score, the names-only prompt selects it at 0.594 (0.588--0.605) against 0.076 (0.053--0.105) for the full prompt and 0.061 for the consensus; the flags-only prompt selects it at 0.456 (0.430--0.482), below the names-only prompt at every seed (12--18 loci to 0, \emph{P}~\(\leq\)~4.9~\ensuremath{\times}~10\textsuperscript{\ensuremath{-}4}). At the loci where the full prompt corrects the evidence ordering, the names-only prompt of the same seed is also correct: at every one of the 9--13 primary loci per seed, and at 25--33 of the 26--34 loci per seed over the whole test set. With the full prompt the top-ranked gene is the first gene of the evidence ordering at 249--257 of 296 loci (84--87\%); without the ordering it coincides with that gene at 140--141 (47--48\%), and the first gene shown is chosen at 67--84 of 296 (23--28\%, against 27\% expected if position decided).

The permuted prompt is a stress test of anchoring under contradictory inputs, not a measure of the evidence content: the text, order, and scores are those of the evaluated prompt, and only the symbols move. With each gene’s symbol attached to another gene’s evidence row, the top-ranked gene is the first-shown row, now carrying another gene’s symbol with the highest score, at 231--238 of 296 loci (78--80\%), and the gene whose own evidence had been shown first at 15--24 (5--8\%). Accuracy follows the row the ordering puts first. On the primary split Precision@1 falls to 0.284 (0.267--0.311), below the full prompt of the same seed (85--90 loci to 17--19, \emph{P}~\(\leq\)~2.4~\ensuremath{\times}~10\(^{- 11}\)), the names-only prompt (82--87 to 6--7, \emph{P}~\(\leq\)~2.5~\ensuremath{\times}~10\(^{- 17}\)) and the evidence ordering (0.711; 82--88 to 25--28, \emph{P}~\(\leq\)~2.6~\ensuremath{\times}~10\(^{- 7}\)); on the 182 naturally retrieved loci, it falls to 0.337 (0.319--0.357) against 0.621 for the ordering. On the force-added loci, it rises to 0.439 (0.430--0.456) against 0.035 for the ordering, for a mechanical reason: the derangement moves the reference symbol off its placeholder row onto a scored one. Of the variant’s 110--114 correct calls over all loci, 62--69 are loci where that row was the first shown (59 loci have two candidates, where the swap is forced) and 41--48 are loci where the reference gene was selected although its symbol was not on the first row. The total over all 296 loci (0.376 against 0.395 for the ordering) is the sum of these opposite effects, not an equivalence. The model knows these reference genes from the trait and the symbol alone; when a scored ordering is shown it follows the top-scored row at most loci, whichever symbol it carries, and the ordering anchors it to the first candidate (\hyperref[tab:s11]{Supplementary Table~11}; the prompts and the permutation at a worked locus are reproduced in \hyperref[note:9]{Supplementary Note~9}).

Three established comparators answer at different subsets of loci; therefore, each was scored where it made a call and paired against the pipeline on those same loci (\hyperref[tab:s3]{Supplementary Table~3}). Across the loci scored by each comparator, no paired accuracy difference was detected for gene-level association~\citep{deleeuw2015}, transcriptome-wide association16, or combined SNP-to-gene linking~\citep{gazal2022} (\emph{P}~=~0.23, 0.15, and 0.18; exact McNemar, unadjusted, as are those below). Restricted to the naturally retrieved loci at which each answers, the pipeline exceeds all three: 0.818 against MAGMA’s 0.576 (19 loci to 3, \emph{P}~=~8.6~\ensuremath{\times}~10\textsuperscript{\ensuremath{-}4}), 0.808 against S-PrediXcan’s 0.538 (\emph{P}~=~1.3~\ensuremath{\times}~10\ensuremath{-}3), and 0.763 against cS2G’s 0.633 (\emph{P}~=~2.9~\ensuremath{\times}~10\textsuperscript{\ensuremath{-}3}). The pooled comparison conceals this for a reason that runs against our own interest: on force-added loci the ordering reverses, MAGMA reaching 0.244 where the pipeline reaches 0.067, because gene-level methods score genes from variant statistics within gene boundaries and never read the candidate ordering, so they are not penalized when the benchmark buries the gold gene at the bottom of it.

On the 88 replicate loci from independent cohorts, which share a reference gene and trait with training and carry no force-added gene, the pipeline’s consensus call is correct at 50 (0.568; 0.735 of the 68 at which a reference gene is among the candidates), against 0.534 for the evidence ordering (3 loci to 0; \emph{P}~=~0.25), 0.398 for nearest-gene assignment (17 to 2; \emph{P}~=~7.3~\ensuremath{\times}~10\ensuremath{-}4), and 0.545 for Open Targets L2G, which predicts at all 88 (6 to 4; \emph{P}~=~0.75). The model departs from the evidence ordering at 15 of 440 decisions (3.4\%) and changes the consensus top gene at 3 loci; Qwen3-32B departs at none (\hyperref[tab:s9]{Supplementary Table~9}).

The two membership baselines are not evidence-free. They rank a candidate first if it is a reference gene in training (for any trait, or for the same EFO term), and membership alone leaves the leading candidate undetermined at 160 and 199 of the 296 test loci, where the composite evidence score decides. Ranked by training frequency alone, with no evidence, the same lookups reach 86/135 and 74/135 on the primary split, and the pipeline exceeds both (35 to 14, \emph{P}~=~0.0038; 44 to 11, \emph{P}~=~8.7~\ensuremath{\times}~10\textsuperscript{\ensuremath{-}6}). Their parity with the pipeline also depends on train--test overlap. In a post hoc analysis of the 97 primary loci that share no reference gene--EFO pair with training, the pipeline reaches 77/97, against 68/97 for the evidence ordering (9 to 0, \emph{P}~=~0.0039), 75/97 for Open Targets L2G (9 to 7, \emph{P}~=~0.80), 68/97 for gene membership (12 to 3, \emph{P~=~0.035),} and 66/97 for trait membership (11 to 0, \emph{P}~=~0.00098) (\hyperref[tab:s8]{Supplementary Table~8}). The improvement over the evidence ordering persists; the comparison with Open Targets is unchanged. On accuracy alone, the pipeline is competitive with current practice and has not been shown to be superior to it.

Across all 296 held-out loci the ordering changes and every method falls (pipeline 0.486, gene membership 0.611, Open Targets L2G 0.365 with uncovered loci scored as failures), because 114 of those loci have a reference gene added at the bottom of the candidate set by construction. On those loci alone, the pipeline scores 0.061 and Open Targets 0.044 with abstentions scored as failures, while a gradient-boosted model trained on all loci scores 1.000, as does a classifier given only the pattern of missing features. That is the artifact, not a result, and it is why the primary analysis excludes them. Failing to retrieve the reference gene is nonetheless a limitation of the pipeline, not only of the benchmark: the reference gene is among the naturally retrieved candidates at 182 of 296 test loci (61.5\%), and the pipeline’s overall accuracy is the product of that recall and its ranking.

\begin{table*}[!t]
\centering
\caption{\textbf{Open Targets coverage under each estimand.} Accuracy where Open Targets predicts (split by whether the credible set comes from the locus’s own study), accuracy with abstentions scored as failures, and, as a combined method rather than Open Targets alone, accuracy when the first stored candidate is credited where it abstains. Quoting the first alone flatters a method permitted to decline, while quoting the second alone penalizes abstention.}\label{tab:2}
\fontsize{8.5}{10}\selectfont
\setlength{\tabcolsep}{3pt}
\renewcommand{\arraystretch}{1.12}
\begin{tabular}{@{}>{\RaggedRight\arraybackslash}p{\dimexpr0.18000\linewidth-2\tabcolsep\relax}>{\RaggedRight\arraybackslash}p{\dimexpr0.23000\linewidth-2\tabcolsep\relax}>{\RaggedRight\arraybackslash}p{\dimexpr0.07000\linewidth-2\tabcolsep\relax}>{\RaggedRight\arraybackslash}p{\dimexpr0.08000\linewidth-2\tabcolsep\relax}>{\RaggedRight\arraybackslash}p{\dimexpr0.44000\linewidth-2\tabcolsep\relax}@{}}
\toprule
\rowcolor{tablehead}
\textbf{Method} & \textbf{Estimand} & \textbf{\emph{n} loci} & \textbf{P@1} & \textbf{Definition} \\
\midrule
\begin{filecontents*}{tables/table_2.tsv}
Shade	CellA	CellB	CellC	CellD	CellE
	Open Targets L2G	quality-when-available	219	0.493	loci where it returns a prediction
	Open Targets L2G	same-study set	72	0.500	credible set from the locus’s own study
	Open Targets L2G	other-study sets	147	0.490	maximum across other studies’ sets
	Open Targets L2G	quality-when-available	102	0.735	loci where it scores every candidate
	Open Targets L2G	coverage-adjusted	296	0.365	all test loci; 77 abstentions as failures
	OT + fallback	combined method	296	0.422	first stored candidate where OT abstains
	Agentic pipeline	answers everywhere	296	0.486	no abstentions
\end{filecontents*}
\csvreader[separator=tab,head to column names,late after line=\\]{tables/table_2.tsv}{}{\Shade\CellA & \Shade\CellB & \Shade\CellC & \Shade\CellD & \Shade\CellE}
\bottomrule
\end{tabular}
\end{table*}

\subsection{Coverage is a result, not a footnote}\label{sec:coverage-is-a-result-not-a-footnote}

Open Targets returned an L2G prediction for 219 of the 296 held-out loci and scored every candidate at 102 loci. Its accuracy is 0.493 on the loci where it predicts and 0.735 where it scores the entire candidate set. Counting its 77 abstentions as failures (the view of a user who must still decide something at those loci) gives 0.365 across all 296 (Table~\ref{tab:2}). Crediting Open Targets with the first stored candidate at the 77 uncovered loci (correct at 17 of them) gives 0.422; that number describes a combined method, not Open Targets, and is shown only labelled as such. Only 72 of the 219 predictions came from a credible set of the locus’s own study; the other 147 took the maximum score across other studies’ credible sets for the same variant, which is a cross-study rather than a query-trait-specific comparison (accuracy 0.500 and 0.490).

Both numbers are legitimate and answer different questions: quality when a prediction is available, against what is delivered in practice. Reporting only the first flatters any method permitted to decline, and reporting only the second penalizes abstention, which may be the correct behavior; a missing record does not by itself show calibrated abstention. We provide both, and note that the pipeline returns a ranked answer with supporting evidence at every locus, including the 77 where Open Targets does not. The worked example in Fig.~\ref{fig:1} is one of them.

\subsection{The contribution is selective, high-precision intervention}\label{sec:the-contribution-is-selective-high-precision-intervention}

Candidates reach the model in evidence order, showing the composite score. The measurable question is therefore not overall accuracy but when the model departs from that ordering and whether it is right to.

\begin{figure*}[!p]
\centering
\includegraphics[width=0.94\textwidth,height=0.50\textheight,keepaspectratio]{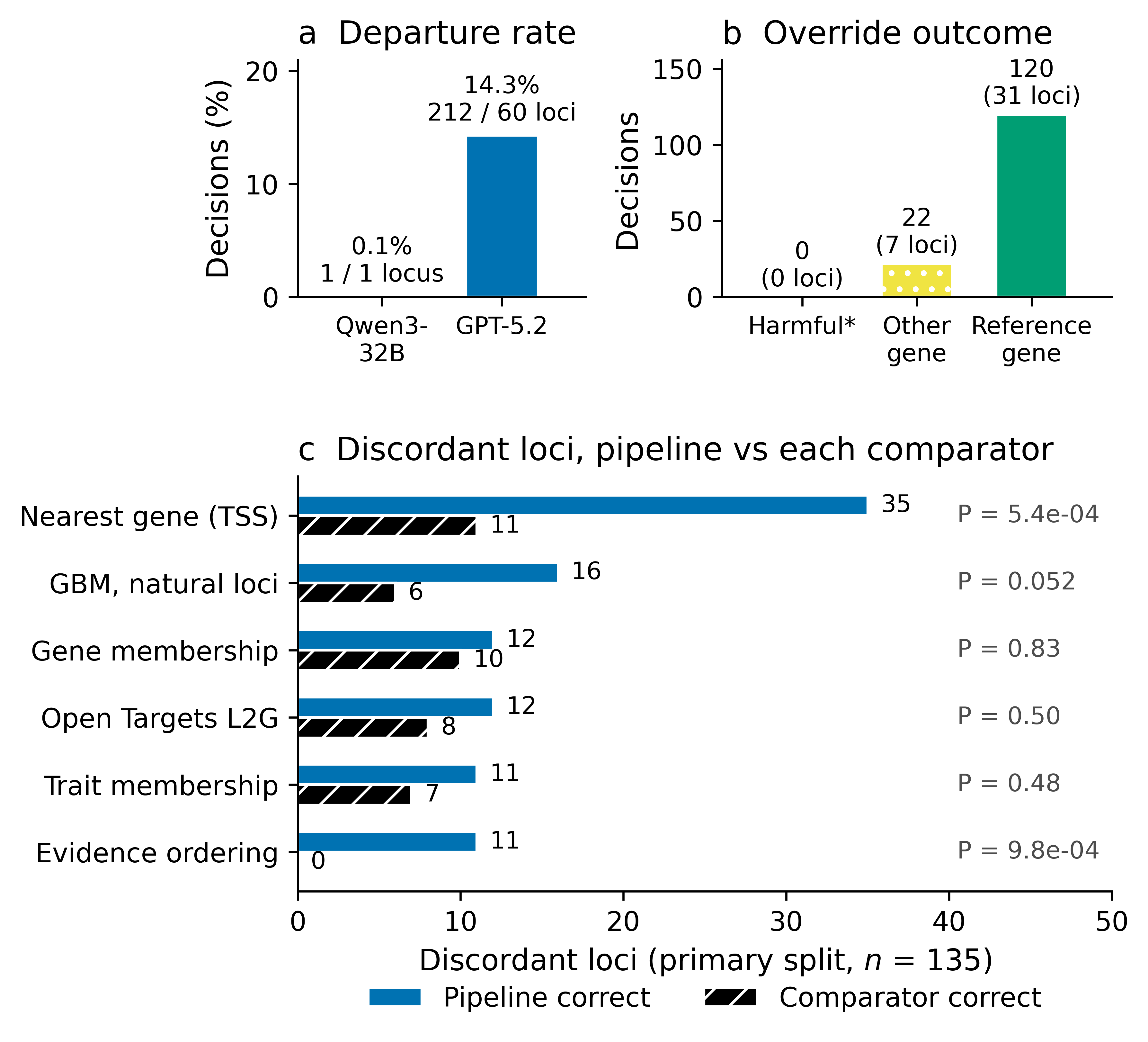}
\caption{\textbf{The pipeline’s contribution is a rare, high-precision intervention rather than a uniformly better ranking.} \textbf{a}, Frequency of each backbone departing from the order in which candidates were presented over 1,480 locus--seed decisions per backbone (296 held-out loci \ensuremath{\times} 5 seeds). GPT-5.2 departed at 212 decisions (14.3\%); Qwen3-32B departed once (0.1\%), and its Precision@1 was consequently identical to the composite evidence ordering in every seed (0.395 in both cases, all five seeds; at that one locus, neither the initial nor the promoted gene was the reference gene). Bars represent counts over decisions, not loci; the 212 GPT-5.2 departures occur at 60 loci. \textbf{b}, Outcome of the 142 departures (37 loci) made by GPT-5.2 on loci whose reference gene was retrieved naturally, that is, excluding loci where the reference gene had to be added to the candidate set. A total of 120 promoted a reference gene (85\%; 31 loci), and 22 did not (7 loci; one locus appeared in both counts because its seeds were split between the two outcomes). *Harmful count departures that displaced a candidate that was already a reference gene; its denominator was all 212 departures, including those on force-added loci, and none occurred. \textbf{c}, Paired outcomes against each comparator on the primary split (135 natural loci at which Open Targets also returned a prediction). The bars show discordant loci only (those where exactly one of the two methods is correct), because concordant loci carry no information in a paired test. The nearest gene was determined by the GENCODE TSS distance. P values are two-sided exact McNemar tests on discordant counts, unadjusted. The full results across all splits are presented in \hyperref[tab:s1]{Supplementary Table~1}.}\label{fig:2}
\end{figure*}

Across 1,480 locus--seed decisions (296 loci \ensuremath{\times} 5 seeds) the pipeline keeps the given top candidate at 86\% of them (Fig.~\ref{fig:2}a); its 212 departures occur at 60 distinct loci. On loci without a force-added reference gene, it departs at 142 decisions (37 loci), and 120 of those (31 loci) select a reference gene (85\% of decisions; Fig.~\ref{fig:2}b); the other 22 (7 loci, one of which is also among the 31 because its seeds split) do not. Seeds of one locus are not independent; therefore, the locus counts are the relevant sample sizes. No departure replaced a top candidate that was already a reference gene. That is useful to report but is not evidence of zero risk: with 117 loci whose initial top candidate was correct, zero events are compatible with a displacement rate of up to 2.5\% per locus (one-sided 95\% bound, assuming independent loci). The five seeds are combined into one gene per locus by a plurality vote that does not use the reference labels (Methods); it changes the initial top gene at 41 of the 296 held-out loci, and the seeds disagree on the top gene at 35.

The contrast with an open-weights backbone sharpens the point. Qwen3-32B departs from the given order once in 1,480 decisions at a force-added locus where neither the initial top candidate nor the promoted gene was the reference gene; therefore, its Precision@1 is identical to the evidence ordering in every seed (\hyperref[tab:s1]{Supplementary Table~1}). Hence, a scaffold of this kind can score respectably while contributing nothing of its own, which is an argument for reporting the departure rate as a routine diagnostic whenever an agent is placed on top of a ranked candidate list.

\subsection{Phenotype annotations support the rankings and the reference genes}\label{sec:phenotype-annotations-support-the-rankings-and-the-reference-genes}

\begin{figure*}[!t]
\centering
\includegraphics[width=0.78\textwidth,height=0.50\textheight,keepaspectratio]{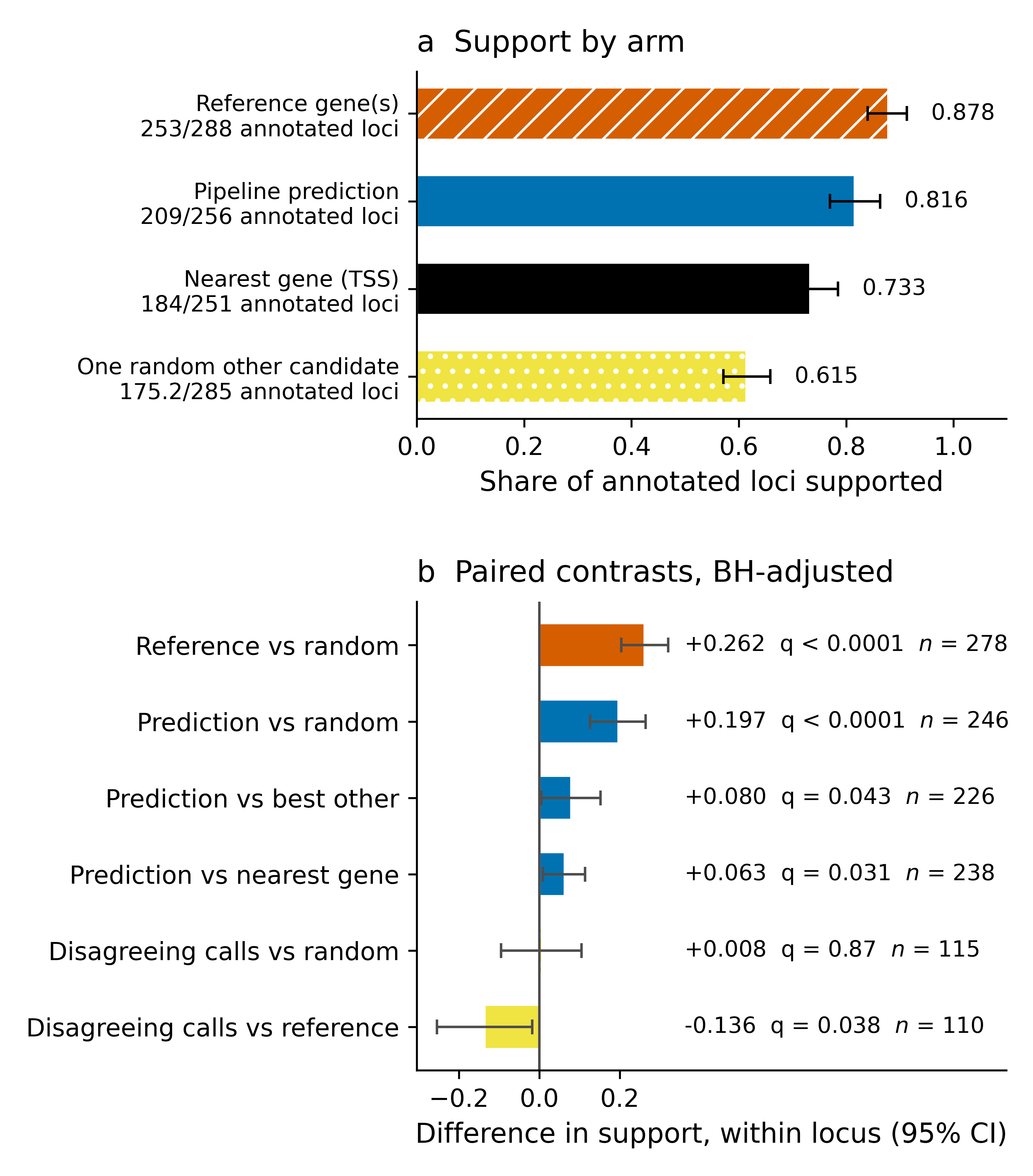}
\caption{\textbf{The pipeline’s calls and curated reference genes carry trait-matched phenotype annotation more often than other candidates at the locus.} Support is assessed against mouse mutant phenotypes (MGI, all allele classes) and human Mendelian disease phenotypes (HPO), with trait and phenotype terms mapped to 18 organ-system categories by versioned rules. These annotations are not GWAS associations but are not independent of the literature either (Results). Predictions are the consensus GPT-5.2 gene per locus. A gene is supported at a locus when any of its phenotype categories meets any of the trait’s. \textbf{a}, Support by arm, shown as supported/annotated loci out of 295 eligible loci, with 95\% bootstrap intervals over loci; the random-candidate arm is the expected support of one uniformly drawn other candidate, so its count is fractional. Because mouse mutants are pleiotropic (an annotated gene carries a mean of 6.8 of the 18 categories), absolute rates are not interpretable on their own, and only within-locus differences are. \textbf{b}, Within-locus paired contrasts. Bars are the difference in support between the two arms at the same locus, with 95\% intervals from a bootstrap over clusters of loci sharing a reference gene; n is the number of loci where both arms have an annotated gene. q values are \mbox{Benjamini--Hochberg} adjusted across the 11 contrasts for this source (exact McNemar or sign-flip permutation tests); solid fill marks q < 0.05 and hatched fill q \ensuremath{\geq} 0.05. All contrasts, by source, are in \hyperref[tab:s5]{Supplementary Table~5}. Seed 20261017.}\label{fig:3}
\end{figure*}

Because accuracy against a curated reference set is a limited test, we compared the rankings with mouse mutant phenotypes (MGI~\citep{baldarelli2024}, all allele classes; 77\% of candidate genes annotated) and human Mendelian disease phenotypes (HPO~\citep{gargano2024}; 35\%; 78\% with either). These are not GWAS associations, but they are not independent of the literature. HPO records human disease relationships, and we have not established whether either resource is absent from the reference curation or from the language model’s pretraining. Trait and phenotype terms were mapped to 18 organ-system categories using versioned rules built from the trait list and ontologies~\citep{gargano2024,smith2009mp} (Methods). The analysis uses the pipeline’s own primary predictions, the consensus GPT-5.2 gene at each locus.

Mouse mutants are highly pleiotropic, with a candidate gene carrying about seven of the 18 categories on average, so absolute support rates are uninformative and only differences between genes at the same locus are interpreted. Of 295 loci with a mappable trait, the prediction was annotated at 256 and supported at 209. Within a locus, reference genes carry trait-matched annotation more often than one randomly drawn other candidate (difference +0.262, 95\% CI +0.204 to +0.320; 278 loci; \emph{q}~<~10\textsuperscript{\ensuremath{-}4}), which is a positive control for the assay rather than a validation of the reference set. The pipeline’s predictions likewise exceed a randomly drawn other candidate (+0.197, +0.126 to +0.264; 246 loci; \emph{q}~<~10\textsuperscript{\ensuremath{-}4}), nearest-gene assignment (+0.063, +0.009 to +0.113; 238 loci; \emph{q}~=~0.031) and the best-supported other non-reference candidate (+0.080, +0.005 to +0.152; 226 loci; \emph{q}~=~0.043) (Fig.~\ref{fig:3}). The advantage persists after adjusting for the number of phenotype categories each gene carries, in a conditional logistic model within locus (log odds ratio 1.16, 95\% CI 0.64--1.69). Most of it comes from predictions that are reference genes (+0.363 over a random candidate, 131 loci).

At the 115 eligible loci where the prediction contradicts the reference, it is not better supported than a randomly drawn other candidate (+0.008, \ensuremath{-}0.095 to +0.105; \emph{q}~=~0.87) and is less often supported than the reference genes it contradicts (\ensuremath{-}0.136, \ensuremath{-}0.255 to \ensuremath{-}0.018; \emph{q}~=~0.038; 110 loci, those with an annotated reference gene). Against a background that excludes the reference genes, the difference is +0.122 (\emph{q}~=~0.038), but that background is defined by the labels and removes the best-supported genes; therefore, we do not rely on it. We detected no additional support for the disagreements. This does not show that none of them reflects unrecognized biology: a significant positive control does not establish power in this smaller subset.

\subsection{An open, deployable implementation}\label{sec:an-open-deployable-implementation}

\begin{figure*}[!t]
\centering
\includegraphics[width=0.96\textwidth,height=0.49\textheight,keepaspectratio]{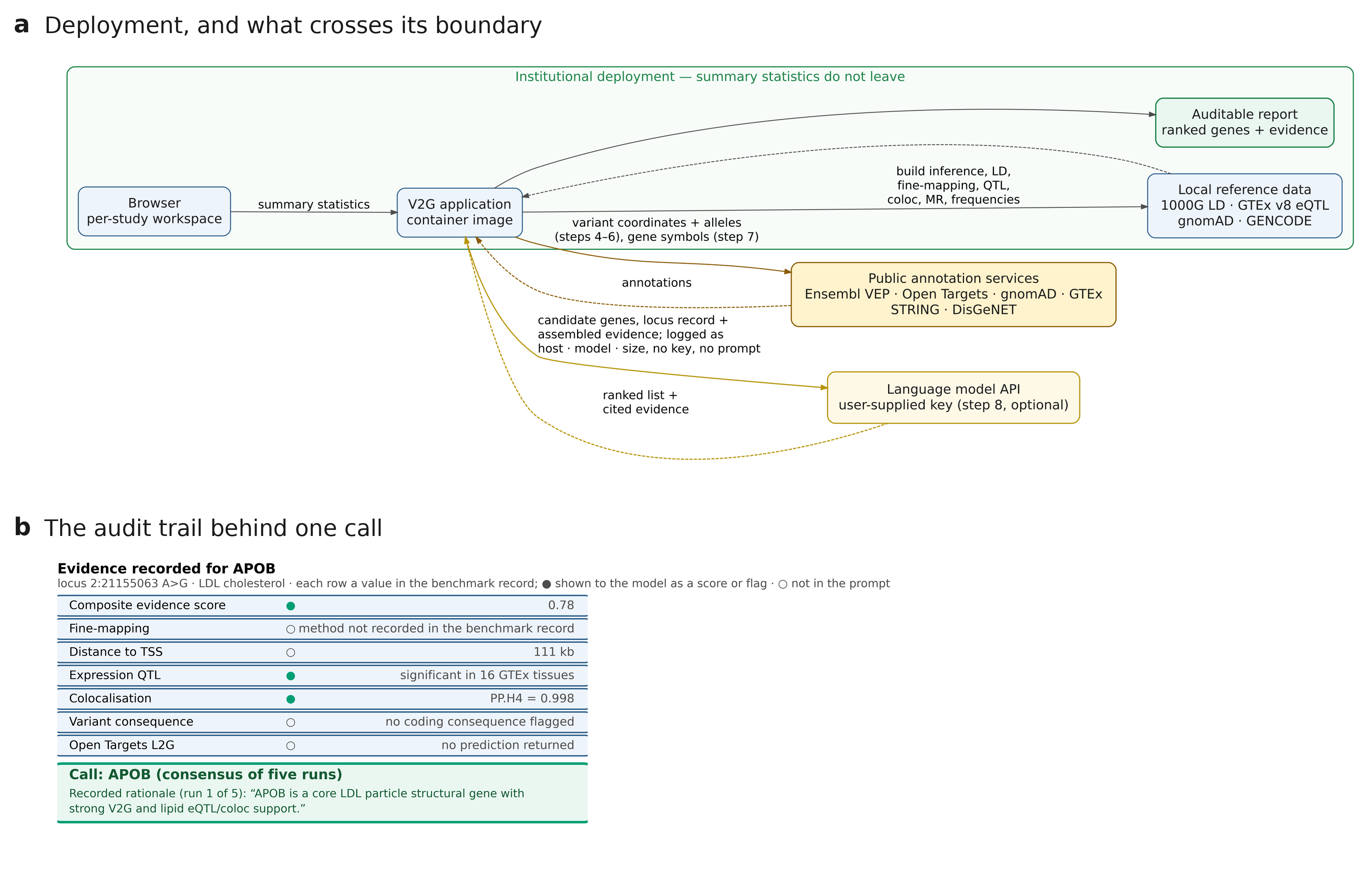}
\caption{\textbf{The deployed implementation and what a user receives.} The figure describes the application; the benchmark runs evaluated in this study used a narrower prompt (Methods, Reasoning, and output); therefore, panel a is not a description of the evaluated configuration. \textbf{a}, Deployment boundary. The application is distributed as a container image and runs inside the user’s own institution. The summary statistics are read inside the container and never leave it. Build inference, linkage disequilibrium, fine-mapping, expression-QTL retrieval, colocalization, Mendelian randomization, and reporting execute against local reference data. Annotation, the Open Targets, gnomAD, and GTEx lookups, and drug-target scoring send variant coordinates and alleles or gene symbols to public services, and the GWAS Catalog browser sends trait and study identifiers; the optional reasoning step sends the candidate genes, locus record, and assembled per-gene evidence to a language-model endpoint with a key the user supplies, and each such request is logged before it is sent as host, model, message count, and payload size, with no key and no prompt text. \textbf{b}, The audit trail behind a single call, shown for the locus of Fig.~\ref{fig:1}. The called gene is the consensus prediction of five GPT-5.2 runs; each row is that gene’s value in the released benchmark record, and the rationale is the text stored for it in the first run, quoted verbatim. The evaluated prompt held only the composite score and the eQTL and colocalisation flags for this gene (\hyperref[note:7]{Supplementary Note~7}); the panel marks which rows were shown. The rationale’s ‘lipid eQTL’ is an inference from the trait name, not a retrieved tissue: no tissue was supplied. A field the record does not hold is printed as not recorded rather than inferred: the fine-mapping method is not stored per locus, no coding consequence was flagged for this gene (which does not by itself show the variant is non-coding), and Open Targets returned no prediction. The colocalization value is computed from incomplete regional inputs (Methods). This schematic of the interface is generated from the recorded benchmark data.}\label{fig:4}
\end{figure*}

The pipeline is released as a multi-user web application with per-study workspaces, interactive visualization, and exportable reports in which each prioritization is linked to its evidence (Fig.~\ref{fig:4}), together with the frozen benchmark, its replicate-cohort extension, the cached agent runs, and the evaluation code, archived as a single citable release. Build inference, linkage disequilibrium, fine-mapping, expression-QTL retrieval, colocalization, Mendelian randomization, and reporting run against local reference data, and the summary statistics never leave the deployment. Three stages do reach public services: variant annotation (Ensembl VEP~\citep{mclaren2016}), the Open Targets~\citep{buniello2025}, gnomAD and GTEx lookups, and drug-target scoring (Open Targets, STRING~\citep{szklarczyk2023}, DisGeNET~\citep{pinero2020}) send variant coordinates and alleles or gene symbols, the optional GWAS Catalog browser sends trait and study identifiers, and the reasoning step sends the candidate genes and their assembled evidence to a model endpoint with a key the user supplies. The application also runs with networking disabled, in which case the first three stages complete and the later ones report the sources they could not reach rather than omit them silently.

A demonstration container image accompanies the release (Methods; \hyperref[note:5]{Supplementary Note~5}). It carries the application and reference-data slices for two loci, the \emph{APOB} locus on chromosome 2 and the 8q24 prostate cancer locus on chromosome 8, with four example inputs, two cohorts per trait (LDL cholesterol from the 2021 and 2013 Global Lipids Genetics Consortium analyses~\citep{graham2021,willer2013}; prostate cancer from a UK Biobank analysis~\citep{harris2024} and the PRACTICAL consortium~\citep{schumacher2018}), one of which is supplied in GRCh37 so that build inference and coordinate conversion can be exercised; the pipeline runs end to end on them without any download. Every saved study records what ran rather than what the interface offered: the build detected and the number of variants converted, the LD panel, the fine-mapping method at each locus and the sample size it used, and which model at which host answered. The image ships with the memory component enabled (EviMem, with its verified claim store CAVE; \hyperref[note:5]{Supplementary Note~5}), so a second cohort of a trait whose locus has already been reasoned about can be ranked from the stored anchor without a model call; this is labelled in the interface and the report, is not part of the evaluation reported here, and can be switched off. A headless script asserts 167 checks on the recorded artifacts of both demonstration studies inside the image; it was run on arm64 and x86-64 machines from the documented commands, by us, which is an installation and run check on a second architecture rather than an independent reproduction.

\FloatBarrier

\section{Discussion}\label{sec:discussion}

On the primary analysis, reranking assembled evidence with a language model improves on the evidence ordering it is given and on nearest-gene assignment; on the primary analysis we detected no difference from Open Targets L2G or the membership lookups, and against gene-level, transcriptome-wide and SNP-to-gene methods it is ahead only on the naturally retrieved loci at which each answers. We think the more useful finding is how it achieves that. It defers to the evidence ordering at 86\% of decisions; on loci where the reference gene was naturally retrieved it departs at 16\% of decisions (142 of 910), and those departures select a reference gene at 85\% of decisions (31 of 37 loci). No displacement of a correct call was observed. The value is not a uniformly better ranking but a small number of corrections where the aggregate score is misleading, as at the LDL locus in Fig.~\ref{fig:1}, where the composite score ranks on proximity and the gene that resolves the locus is the one whose biology is known rather than the one that is nearest. The prompt ablations say where those corrections come from: given the symbols and the trait alone the model is right more often than with the evidence, so the corrections are recall of the gene’s known biology, and the evidence ordering, as presented, mostly keeps the model from applying it.

General-purpose biomedical agents now exist, and Biomni~\citep{huang2025biomni} lists GWAS causal-gene detection among the workflows on which it is evaluated. We have not compared against it, and the stored runs evaluated here make one language-model call per locus over assembled evidence, so they establish reranking performance, not adaptive tool selection, reliable tool execution or explanations supported by the retrieved evidence. The pipeline provides recorded evidence retrieval and gene rankings that can be inspected locus by locus. The ablations measure how rankings change when gene names or the presented evidence are altered; on this benchmark, they do not show an accuracy gain from evidence integration. We have not measured whether each statement in a rationale is supported by a recorded retrieval.

The ablations also set the limit of the claim. Presenting candidates in evidence order with the score attached anchors the model: with the full prompt its top gene is the first candidate at 84--87\% of loci, without the ordering at 47--48\%, and the model scores higher without the evidence than with it, at every seed and also on the 97 primary loci whose reference gene and trait never co-occur in the training set. The retrieved flags do not help: on the primary split, they change nothing (0.862 against 0.864), and on the force-added loci, they lower accuracy (0.456 against 0.594). Two classifiers restricted to the prompt’s four fields score below the evidence ordering (Results), so the gain over the ordering was not recovered by classifiers from what was shown; it is recall of reference genes that the model names from the symbol and the trait alone, including 59\% of the reference genes the pipeline never retrieved, and at every primary locus where the full prompt corrects the ordering the names-only prompt is also correct. The permuted prompt says what the model does when a scored ordering and its own knowledge disagree: it takes the top-scored row at 78--80\% of loci and the gene whose own evidence was first at 5--8\%. Contradictory inputs could themselves push a model to defer to the score, but the unpermuted prompt, with nothing contradictory in it, shows the same deference (the first candidate kept at 84--87\% of loci), so the permutation exposes a habit rather than creating one. The variants are not a factorial design: the two reduced prompts share a system prompt that no longer announces scores, the permuted prompt keeps the evaluated one (both in \hyperref[note:9]{Supplementary Note~9}), and the primary runs predate the ablations by four months on a deployment whose served version the API does not report. A full-prompt pass repeated in October scored 0.756 on the primary split (102/135) against 0.778--0.807 for the June seeds, and paired with each June seed, it was never correct where that seed was wrong (3--7 loci to 0), so the October output was lower; whether the served model changed or one repetition fell low, the API gives no way to tell. Measured against that October pass, the names-only advantage is wider (7 loci to 21--22, \emph{P}~\(\leq\)~0.013), so neither reading accounts for it. Reasoning over the presented evidence is not what these runs measure; they measure a model that accepts the ordering it is handed and recalls the answer where the ordering leaves room. The open-weights backbone makes the anchoring concrete (it departs once in 1,480 decisions and scores exactly what the evidence ordering scores) and it is the reason we advocate reporting departure rate, a names-only control, and a permuted-evidence control whenever an agent is layered on a ranked list. A scaffold can look competitive while contributing nothing, and an agent can look as if it integrates evidence while recalling the answer.

Two training-membership lookups are not distinguishable from the pipeline on the primary analysis and fall behind it in the post hoc subset without gene--EFO overlap with training. This pattern is not peculiar to this benchmark: evaluation sets drawn from published literature reward recall of what a method may already have seen. None of our overlap measures speaks to what a language model absorbed in pretraining, which is the general form of the contamination problem documented for language-model benchmarks~\citep{sainz2023}. The names-only control does: a model that names the curated reference gene at 0.864 of the primary loci, and at 0.594 of the loci whose reference gene the pipeline never retrieved, from the symbol and the trait alone carries that association without any retrieved evidence, whatever its source in training, and on such a benchmark an agent’s accuracy measures recall before it measures inference.

Coverage deserves more attention than it usually receives in the literature. Open Targets declines to predict at a quarter of our held-out loci, and its reported accuracy differs by 0.13 depending on whether those abstentions are scored. Neither number is incorrect; quoting one without the other is. For a user deciding where to spend experimental effort, a method that answers everywhere with inspectable evidence and a method that answers well three-quarters of the time are not interchangeable, and a single headline accuracy cannot express the difference. The pipeline’s own coverage limit lies upstream of the model: at 114 of the 296 held-out loci (38.5\%), candidate retrieval did not return the reference gene, and the model did not recover it there (Precision@1 0.061 even after the gene is force-added). Answering at every locus is not being correct at every locus, and a user acting on a call at a low-evidence locus bears that risk; the departure rate and the evidence shown with each call are what make it visible.

Phenotypic analysis supports two conclusions and provides no evidence for a third. Curated reference genes carry more trait-matched mouse mutants and Mendelian phenotype annotations than other candidates at their loci, and the pipeline’s calls exceed the nearest-gene assignment on the same annotations. Calls that contradict the reference are not better supported than a random candidate; therefore, we have no evidence that they represent overlooked biology and no evidence that they do not. Organ-system categories provide broad corroboration, not a mechanism. Because the enrichment comes from calls that agree with the curated reference, it corroborates those calls against the nearest-gene and random baselines; it does not show an advantage over a comparator with the same reference-gene recall, such as Open Targets L2G, and the categories are themselves curated from literature that a language model may have seen.

However, several limitations of this study should be considered. The train/test split does not fully separate associations, and we did not re-split by association and retrain the supervised and memory-based comparators; the overlap-free results above are post hoc sensitivity analyses. The replicate cohorts shared their associations with training. We characterized two language-model backbones, and the result depends on which is used; others may differ, particularly where the model’s own biological knowledge is used. We did not compare against general biomedical agents, and the worked example is a familiar association; a less familiar, independently assessed example and an error case would test more. The non-circular curated gold caps are at roughly 1,279 loci; therefore, this benchmark cannot be enlarged indefinitely without reintroducing the circularity it exists to avoid. Cross-ancestry evidence was best powered for the Finnish replicates (\emph{n}~=~35) and thinner for East Asian (\emph{n}~=~19) and South Asian (\emph{n}~=~17); it was not powered for African (\emph{n}~=~8) or Hispanic or Latin American (\emph{n}~=~3) cohorts, and the remaining six replicate loci (Greater Middle Eastern 4, Central Asian 1, European 1) were too few to assess. We fine-mapped by approximate Bayes factor or SuSiE and have not tested sensitivity to the fine-mapping method; FINEMAP~\citep{benner2016} and related approaches may alter which variants, and therefore which candidate genes, enter a locus; and the benchmark candidate sets were built with an LD step that kept the 500 lowest-coordinate matched variants per locus, a rule that can omit the lead or causal variants, at a rate we have not quantified; the released application keeps the 500 variants nearest the lead and always retains the lead. Colocalization in the frozen benchmark was available at 17 of 296 test loci and was computed from incomplete regional inputs (Methods). The functional comparison matches at organ-system resolution, which is coarse, uses mouse mutants of all allele classes, and tests relative enrichment within a locus rather than the mechanism; finer, tissue-specific, and blinded analyses would test more. Reasoning traces are generated texts and should be read as a record of the evidence a model used, not as a mechanistic claim. We report no independent reproduction.

What we would take from this is modest and, we think, useful. On a benchmark that resists recall only in part, an auditable agent is competitive with the best available scores while answering at every locus and showing its working; its measurable contribution is a small number of corrections to the evidence ordering; and its calls are better supported by phenotype annotations than other candidates at the locus, although those annotations are not independent of the literature a model may have absorbed. Whether any method, including ours, reasons its way to a causal gene it could not have remembered remains open, and we have tried to report what we found rather than what would have been a better story.

\section{Methods}\label{sec:methods}

\subsection{The pipeline}\label{sec:the-pipeline}

\textbf{Input and genome build.} The pipeline accepts GWAS summary statistics in tabular form. The genome build is inferred from the coordinates themselves rather than declared by the user: chromosome--position pairs are intersected with a paired site list in which ~1.3~million variants carry both GRCh37 and GRCh38 positions~\citep{hapmap2010}, and the build whose coordinates match more often is taken, with the call reported as confident only when it exceeds 50 matches and a five-fold ratio over the alternative. GRCh37 input is lifted to GRCh38 with the UCSC chain~\citep{kent2002}, strand-aware: for minus-strand mappings the alleles of SNVs are complemented so that the effect allele remains the same physical allele, and minus-strand indels are dropped because their anchor base changes.

\textbf{Sample size.} For fine-mapping methods that require sample size, it is read from the summary statistics, preferring an effective sample size, then case and control counts combined as 4/(1/\emph{n}\textsubscript{cases} + 1/\emph{n}\textsubscript{controls}), then a total; the median of the per-variant values is used, and values that cannot be a sample size are ignored.

\textbf{Loci and linkage disequilibrium.} Variants are first grouped into loci: they are sorted by P-value, and each remaining variant is assigned to the most significant unassigned lead within \ensuremath{\pm}1~Mb on the same chromosome, so each variant belongs to one locus, and loci share no variants; the assignment radius spans up to 2~Mb, and the genomic spans of neighboring loci can still overlap. Variants without coordinates are kept as singletons. LD is then computed per locus from local 1000 Genomes reference panels for five ancestry groups (EUR, AFR, AMR, EAS, and SAS) in both GRCh38 and GRCh37~\citep{tgp2015}. The extraction region is \emph{not} a fixed window around the lead variant: genotypes are loaded from the span of the locus’s own matched reference positions plus a 100~kb flank on each side, with at most 500 variants per locus. The released application keeps the 500 matched positions nearest the lead variant, which is always retained. The benchmark’s candidate sets were built with a rule that kept the 500 lowest-coordinate positions, not the 500 nearest the lead or most significant; it can omit the lead or likely causal variants, and its effect on the benchmark has not been quantified. Regions already computed for an overlapping locus are reused when they contain the query window; therefore, the genotypes loaded may cover more than the requested span, and the returned matrix is restricted to the requested positions. Matrices are cached per locus.

\textbf{Fine-mapping.} Credible sets are derived either by the approximate Bayes factor~\citep{wakefield2009} with prior variance \emph{W} = 0.04 on the scale of the effect estimate (a prior standard deviation of 0.2, applied alike to quantitative traits and log-odds; sensitivity to \emph{W} was not tested) and 95\% credible threshold, or by the summary-statistics form of SuSiE, SuSiE-RSS~\citep{wang2020susie,zou2022}, when an LD matrix is supplied. SuSiE is run with a maximum of \emph{L} = 5 causal signals, 95\% credible-set coverage, a minimum absolute within-set correlation of 0.5 and at most 100 iterations, with the sample size taken from the summary statistics.

\textbf{Evidence retrieval.} For each candidate gene at a locus, the pipeline issues separate, recorded queries for: expression QTLs from a local GTEx v8 store~\citep{gtex2020}; colocalization, with priors \emph{p}\textsubscript{1} = \emph{p}\textsubscript{2} = 1 \ensuremath{\times} 10\textsuperscript{\ensuremath{-}4} and \emph{p}\textsubscript{12} = 1 \ensuremath{\times} 10\textsuperscript{\ensuremath{-}57}; chromatin and distance evidence; variant consequences~\citep{mclaren2016}; per-population allele frequencies from gnomAD genomes~\citep{karczewski2020} for GRCh38 input and 1000 Genomes genotype counts~\citep{tgp2015} for GRCh37; and two-sample Mendelian randomisation~\citep{pierce2013} using the eQTL store as the exposure.

The eQTL store is built from the significant variant--gene association tables from the 49 GTEx v8 tissues, so significance is GTEx’s own per-gene permutation threshold rather than one we impose, and retrieval is by gene rather than by a window around the lead variant. Because the store holds only significant variant--gene pairs, it suits annotation lookup but is not the complete regional input that colocalization assumes~\citep{giambartolomei2014}, and colocalization computed from it, against genome-wide-significant GWAS variants, can overstate shared signal. In the frozen benchmark, colocalization values were computed this way for candidates at 17 of the 296 test loci (the traits with local summary statistics); this may explain the uniform posteriors near 0.99 at the \emph{APOB} locus, and the per-locus inputs are not released. For Mendelian randomization, instruments are the eQTL variants for the exposure gene, filtered on instrument strength at \emph{F} > 10, which removes the weakest instruments but does not guarantee negligible weak-instrument bias~\citep{burgess2011}; palindromic variants are oriented by allele frequency and excluded when either minor allele frequency exceeds 0.42; instruments are not LD-pruned, so correlated instruments enter inverse-variance weighting~\citep{burgess2013} as if independent (a Wald ratio is used for a single instrument). No heterogeneity or pleiotropy diagnostics are applied, and an expression-based estimate alone does not establish mediation by the gene.

\textbf{Reasoning and output.} Candidate genes and their assembled evidence are presented to a language model, which returns a ranked list of the evidence used. Calls are capped at 800 output tokens with three retries on rate limiting. \textbf{Candidates are presented in descending order of the composite evidence score, with that score shown}; this is a property of the implementation that matters for interpretation and is discussed in the main text. In the evaluated benchmark runs, the prompt was narrower than the application’s. For each gene, it carried one line: the composite score (labelled \texttt{v2g}), the flags \texttt{e}\texttt{QTL}\texttt{, }\texttt{coloc}\texttt{,} and \texttt{coding\_\allowbreak{}variant} where set, a field labelled \texttt{L2G} where the record held one (2,125 of 4,517 candidates; in the benchmark version then in use it duplicated the composite score), and two fields, \texttt{CIR} and \texttt{druggability}, that were unpopulated in the benchmark and therefore read 0.00 and 1 for every gene. It held no distances, tissue counts, or posterior values, and no Mendelian randomization estimates (the benchmark holds none). The system prompt asked for the most likely causal therapeutic target with a probability per gene; these probabilities are not constrained to sum to one and are not used as confidence measures here. The stored runs keep the parsed response but not the prompt text; the system prompt and the user-prompt template are reproduced from the runner code in \hyperref[note:7]{Supplementary Note~7}, with the prompt as rendered for the locus of Fig.~\ref{fig:1}.

\textbf{Prompt ablations.} Three further GPT-5.2 variants were run over the 296 held-out loci, three seeds each (2,664 calls), with the same runner, client settings, benchmark tasks, and candidate lists as the primary runs, differing only in the prompt. In the first (\emph{names only}) the prompt carried the trait and the candidate gene symbols in a shuffled order and nothing else. In the second (\emph{flags only, shuffled}) each gene’s line carried the three binary flags (eQTL, colocalization, coding consequence) but no composite score, no L2G value, and neither of the constant CIR and druggability fields, again in shuffled order; for these two the system prompt was changed only so that it no longer announced inputs the prompt did not carry. In the third (\emph{evidence permuted}) the evaluated prompt was left unchanged in text and in order, and the gene symbols were reassigned among the evidence rows of the locus by a derangement, so that no gene kept its own evidence and the gene shown first, with the highest score, was never the gene whose evidence had placed it first. This separates a model that ignores the evidence, whose accuracy would match the names-only prompt, from one that follows it, whose top gene would follow the permuted ordering; because the scores, the order and the framing of the candidates as pre-scored are unchanged, it is a stress test of anchoring under contradictory inputs, not an estimate of the contribution of the evidence content. The shuffle and the derangement are drawn deterministically from the seed and the locus identifier; the shown order, the permutation, the prompt text as sent and the API usage record are stored with each run. A response naming a symbol outside the candidate set is scored as incorrect, as in the primary runs (two of the 2,664 responses). Each run is scored top-1 per locus on the splits of Table~1, and on the 97 primary loci whose reference gene and trait pair is absent from the training reference genes (the overlap-free subset of the sensitivity analysis), and paired by exact McNemar test with the full-prompt run of the same seed, with the five-seed consensus and with the evidence ordering; seeds are repeated calls, not independent loci, and are never pooled into one test. How often the top gene coincides with the first gene of the hidden evidence ordering, with the first gene shown and, for the permuted variant, with the gene whose own evidence was shown first is reported as a diagnostic. Because the primary runs were made in June 2026 and the ablations in October 2026, one further full-prompt pass over the 296 loci was made in October beside the ablations, with the same runner, prompt, client and benchmark tasks, as a control for drift of the served model; it is reported as a single repetition in \hyperref[tab:s11]{Supplementary Table~11}, paired with each June seed and with each variant, and enters no primary comparison (\hyperref[note:9]{Supplementary Note~9} and \hyperref[tab:s11]{Supplementary Table~11}).

\textbf{Implementation and deployment.} The application is a Streamlit web interface over the pipeline library, organized as nine steps (upload and build inference, linkage disequilibrium, fine-mapping, annotation, QTL mapping and colocalization, evidence combination and ranking with optional Mendelian randomization, drug-target scoring, reasoning, and report), each with its own run control and each saved with the study. Provenance is recorded from what ran, never from an interface default: the detected build and conversion counts, the LD panel, the per-locus fine-mapping method and the sample size it used, and the model and host that answered. The reasoning request contains the trait, the candidate genes, the locus record (lead variant; the top posterior inclusion probability and the variant that carries it; whether the lead is in a credible set; the other candidate and the non-coding genes at the locus) and one compressed evidence line per gene (composite score, druggability and direction of effect, expression-QTL tissue count with the top tissue, colocalization posterior and tissue, Mendelian randomization estimate, variant consequence, ClinVar status~\citep{landrum2018}, allele frequency and Pharos target-development level~\citep{kelleher2023}); with memory enabled it also contains recalled anchor templates, each labelled same-trait or cross-trait. Every request is logged before it is sent as host, model, message count, payload size, and local or external destination, with no key and no prompt text. The demonstration image is a Docker container~\citep{boettiger2015} built from a single recipe on Debian Bookworm (Python 3.11.17; Streamlit 1.65.0 pinned; the full package list frozen from the built image and deposited) for linux/amd64 and linux/arm64. It bundles the 1000 Genomes GRCh38 genotypes of all five superpopulations for the two demonstration windows (2,503 samples), the GTEx v8 significant variant--gene pairs within them, the complete GENCODE v38 TSS table, the paired site list used for build inference, and the GWAS Catalog trait database with its EFO terms~\citep{cerezo2025,malone2010}. Uploads and the other input types are disabled in that image because no reference data exists outside the two windows; each limit is switched off explicitly and is lifted by one environment variable once the full reference set is present. A verification script drives both demonstration studies headlessly through parse, build inference, conversion, LD, fine-mapping, annotation, ancestry frequencies, and scoring, and then the memory write path, and asserts on what was recorded (167 checks); a second checks the host architecture, the imports, the bundled data, and the service health (36 checks).

\subsection{Benchmark}\label{sec:benchmark}

\textbf{Loci.} 844 fine-mapped loci across 180 traits and 175 EFO terms, drawn from a curated non-circular gold standard over the GWAS Catalog~\citep{cerezo2025}, split 548 train / 296 test. The split was intended to hold out gene--trait associations. After splitting, overlap was measured by treating a test locus as overlapping when any of its reference genes matches: 87 test loci share a reference gene and exact trait name with training (so 209 do not), 99 share a reference gene and EFO term (197 do not), 154 share a reference gene with any training locus, 216 share the EFO term and 7 share the lead variant. Ontology synonyms and parent--child relations between EFO and MONDO terms are not resolved; therefore, the gene--EFO count is a lower bound on association overlap. A further 88 replicate loci from independent cohorts are reserved for transfer; by the benchmark’s \texttt{target\_\allowbreak{}ancestry} field these are Finnish 35, East Asian 19, South Asian 17, African 8 (7 African American or Afro-Caribbean, 1 African unspecified), Greater Middle Eastern 4, Hispanic or Latin American 3, Central Asian 1 and European 1. The coarse \texttt{target\_\allowbreak{}pop} column in the same file collapses all but the Finnish, African, and admixed loci into a single \texttt{eas} bucket of 42; \texttt{target\_\allowbreak{}ancestry} is the correct field. Candidate sets contain a median of four genes.

\textbf{Candidate-set record.} The released benchmark applies three construction rules, documented in full in \hyperref[note:1]{Supplementary Note~1}: (i) reference genes that the evidence pipeline did not retrieve are added to the candidate set with an explicit \texttt{force\_\allowbreak{}added} flag and no composite score (309 rows, 301 loci across all splits; 114 loci in the held-out set), because a placeholder score would label such genes by their own missingness; (ii) genuine Open Targets L2G~\citep{mountjoy2021,ochoa2023} is stored in its own column, \texttt{l2g\_\allowbreak{}ot}, with \texttt{l2g\_\allowbreak{}ot\_\allowbreak{}source} recording, per locus, whether the credible set came from the locus’s own study (the field is set at locus level, so it is also populated for candidates whose \texttt{l2g\_\allowbreak{}ot} is null), and is never folded into the composite score; (iii) \texttt{v2g\_\allowbreak{}origin} records which of three quantities each composite score is (2,125 copied from L2G, 2,083 distance-derived, 309 force-added rows with no evidence-derived score). The benchmark is frozen and checksummed. The stored agent runs were made on an earlier build of the same candidate sets, identical in membership and ordering; the values its prompts carried are described in \hyperref[note:7]{Supplementary Note~7} (see also Reasoning and output), and every result is scored on the released benchmark.

\textbf{Splits and estimands.} Force-added and natural loci are reported separately throughout, with the pooled result over all 296 loci given alongside the results. The primary analysis is the 135 natural loci at which Open Targets also returns an L2G prediction; the results are conditional on successful candidate retrieval and Open Targets coverage. For any method that may decline to answer, two estimands are reported: \emph{quality-when-available}, conditioned on the loci where it predicts, and \emph{coverage-adjusted}, in which abstention is scored as a failure across all 296 loci.

\subsection{Comparators}\label{sec:comparators}

\textbf{Open Targets L2G.} Scores were fetched from the Open Targets Platform GraphQL API (v4)~\citep{ochoa2023}, querying credible sets by variant and taking the L2G predictions~\citep{mountjoy2021} attached to each. Where a credible set from the task’s own study was available, it was used (72 of 296 loci); otherwise, the maximum score per gene across other GWAS credible sets for that variant was taken (147 loci), a cross-study comparison rather than query-trait-specific L2G. Candidates with no prediction rank last. Open Targets returned a prediction for 219 loci and scored every candidate at 102; at the other 77 no prediction was returned. Three estimands are reported: accuracy on covered loci, accuracy with uncovered loci scored as failures, and (labelled as a combined method) Open Targets with the first stored candidate as a fallback at uncovered loci. The query requested the first 50 credible sets per variant without further pages, the stored outputs do not distinguish a failed request from an empty result, and the API version does not identify the data or model release; retrieval and matching are detailed in \hyperref[note:2]{Supplementary Note~2}.

\textbf{Memorisation baselines.} It is defined only for the training loci. \emph{Gene membership with evidence-score tie-breaking} ranks first any candidate that is a reference gene at some training locus, and \emph{trait membership with evidence-score tie-breaking} any candidate that is a reference gene for the same EFO term in training; membership carries weight 106, and the composite score orders candidates within a class, which decides the leading candidate at 160 and 199 of the 296 test loci. Two evidence-free controls rank by \emph{frequency} alone (the number of training loci at which the gene, or the (gene, EFO term) pair, is a reference), with ties broken alphabetically (137 and 195 loci).

\textbf{Nearest gene.} The candidate whose transcription start site is closest to the lead variant from a GENCODE v38 TSS table~\citep{frankish2021}; ties are broken deterministically. The benchmark’s stored distance field is not used: it is present at only 97 of the 296 test loci, and ranking by it falls back to the first stored candidate (the evidence-ordering top) at the other 199 (0.696 on the primary split, against 0.615 by TSS).

\textbf{Gene-level, transcriptome-wide, and SNP-to-gene comparators.} MAGMA~\citep{deleeuw2015}, S-PrediXcan16, and cS2G~\citep{gazal2022} were each run once on this benchmark, and their calls were reused without recomputation; their per-locus calls were re-scored onto the splits defined above so that they were comparable with everything else reported. Each answers at a different subset of loci (MAGMA at 111, S-PrediXcan at 83, and cS2G at 236, where it makes a call within \ensuremath{\pm}2~kb), so each is scored only where it answers and is paired against the pipeline on exactly those loci. For cS2G, loci where no call is made are excluded rather than scored: the stored per-locus file ranks those loci by the composite evidence ordering as a fallback, so scoring them would measure the fallback and not cS2G. The one difference is invisible in the aggregate (the fallback scores 0.433 against cS2G’s own 0.432), which is precisely why it is stated here. PoPS~\citep{weeks2023} is not among the comparators: it needs genome-wide MAGMA gene statistics for every study and a gene-feature matrix, and MAGMA could be run only for the studies with usable full summary statistics (111 loci); the benchmark’s \texttt{pops\_\allowbreak{}subscore} column was never populated and was dropped in v2.

\textbf{Supervised model.} Gradient-boosted trees (XGBoost~\citep{chen2016xgboost} 3.2.0, scikit-learn 1.8.0, pandas 2.3.3, NumPy 2.1.0, Python 3.11.14, the reproduction archive’s pinned environment; 300 estimators, maximum depth 3, learning rate 0.05, subsample 0.8, column subsample 0.8, random state 0), with median imputation fitted to training data only. Because removing the 0.1 placeholder did not remove the label signal (a force-added reference gene now carries missing values instead, which a tree learns equally well), the reported supervised figure is a \textbf{gradient-boosted model trained on naturally retrieved loci}, evaluated on every split. This restriction removes the force-added missingness signal from training and does not address the train--test association overlap. Results depend on the library versions; a rerun with newer versions changed this model from 97/135 to 98/135 on the primary split. The unrestricted variant and a \emph{missingness-only control}, trained on the pattern of missing features with every value withheld, are reported alongside as diagnostics; both reached 1.000 on force-added loci.

\subsection{Agent runs}\label{sec:agent-runs}

Two backbones: Qwen3-32B served through Dashscope, and GPT-5.2 through Azure OpenAI, each run with five seeds over the 296 held-out loci (1,480 locus--seed decisions per backbone); the five seeds are five repetitions of otherwise identical calls (no sampling seed is passed to the API; the seed index governs only the bandit configurations, which are not evaluated here). Qwen3-32B was called at temperature 0.1 with thinking disabled. For GPT-5.2, the client sends neither a temperature nor a reasoning-effort parameter; therefore, the API defaults are applied; the per-call usage records stored with seven of the nine prompt-ablation runs (2,072 calls, made through the same code path) report no reasoning tokens and a completion cap of 800 tokens that no call reached. One gene per locus was selected without reference to the labels: each seed votes for its top-ranked gene, the gene with the most votes is taken, and ties go to the better mean rank across seeds, then to the evidence ordering (the selected genes are released). Correctness was scored after selection. All ten primary runs contain 296 predictions with no recorded errors. Every number in Table~1 and every paired test was scored from these cached runs of the frozen evaluation; the only other model runs were the prompt ablations described below (three variants, three seeds each), which are reported separately and enter no primary comparison.

\textbf{Departure analysis.} For each locus and seed, the top-ranked gene is compared with the candidate holding the highest composite score, the first gene presented. A \emph{departure} occurs when the genes differ. Departures are reported separately for loci with and without a force-added reference gene, because on the former, the gene is buried by construction, and any apparent gain would be an artifact; both decision and locus counts are given. For each departure, we record whether the newly promoted gene is the gold gene and whether the displaced candidate was (\hyperref[tab:s4]{Supplementary Table~4}).

\subsection{Orthogonal functional support}\label{sec:orthogonal-functional-support}

\textbf{Sources.} Mouse mutant phenotypes from MGI~\citep{baldarelli2024} (the human--mouse phenotype report, with human symbols mapped to MP terms and orthology pre-resolved; spontaneous, induced and engineered alleles alike, as no allele-class filter is applied) and human Mendelian phenotypes from the HPO~\citep{gargano2024} (gene-to-phenotype annotations), with the MP~\citep{smith2009mp} and HP~\citep{gargano2024} ontologies for roll-up. 77\% of candidate genes carry mouse mutant annotations, 35\% human Mendelian, and 78\% either. Downloaded 2026-10-01; provenance and versions recorded alongside the files. Deliberately excluded: ENCODE-rE2G and Open Targets perturbation atlases, because the pipeline already consumes regulatory evidence and Open Targets is a comparator, so support from either would be circular.

\textbf{Category mapping.} Trait names and phenotype terms were mapped to 18 organ system categories, as detailed in \hyperref[note:4]{Supplementary Note~4}. Trait categories come from keyword rules over the trait list, recording which rule fired. Phenotype categories come from rolling every annotated term up to the top-level system terms of its ontology (the direct children of MP:0000001 and HP:0000118, read from the ontology files) and mapping those to the same vocabulary, with categories carrying no organ-system information (cellular, embryonic, prenatal, constitutional, “normal phenotype”) dropped explicitly. Both maps were built from the trait list and the ontologies alone, without reference to predictions or reference genes, and were released as dated CSVs (2026-10-01; byte-identical in every later copy); the file dates alone cannot show that the rules were fixed before any prediction was examined.

\textbf{Test.} A gene is \emph{supported} for a locus when any of its phenotype categories meets any of the trait’s. Because mouse mutants are pleiotropic (across all annotated genes a gene carries a median of five and a mean of 6.8 of the 18 categories, and across the benchmark’s candidate genes specifically a median of seven and a mean of 7.8), absolute support rates are uninformative, and only differences between genes at the same locus are interpreted. Predictions are the consensus GPT-5.2 gene per locus (Agent runs). A locus is \emph{eligible} when its trait maps to a category; a gene is \emph{annotated} when it has any category, and \emph{supported} as above; all three counts are reported per arm. Arms: the prediction; the reference genes (a locus counts as supported if any is; a positive control for the assay); the nearest gene by TSS; \emph{one random other candidate}, whose value is the exact expectation of a single uniform draw from the candidates other than the prediction (the mean support among them), and which does not depend on the labels; the same excluding reference genes, which does; and the best-supported other non-reference candidate, a maximum that grows with candidate number. Only annotated genes contribute, and a locus enters a contrast when both arms have one. Binary contrasts use the exact McNemar test; contrasts against the random-candidate arms use a paired sign-flip permutation test (100,000 permutations). Effects carry 95\% bootstrap intervals~\citep{efron1979} (10,000 resamples) over loci and over clusters of loci sharing a reference gene, and \emph{P}-values are \mbox{Benjamini--Hochberg} adjusted~\citep{benjamini1995}, unrounded, within each annotation source (any, MGI, HPO; 11 contrasts each); seed 20261017. Annotation burden is controlled by a conditional logistic regression within locus of support on being the prediction and on the gene’s number of categories. All analyses use the released benchmark, category maps and cached predictions. All contrasts are in \hyperref[tab:s5]{Supplementary Table~5}.

\subsection{Statistics}\label{sec:statistics}

Paired comparisons between methods used the exact McNemar test~\citep{mcnemar1947} on discordant loci, the paired form being appropriate because all methods identified identical candidate sets. Where several comparisons form a family, \emph{P}-values are Benjamini--Hochberg corrected~\citep{benjamini1995} and reported as \emph{q}. Accuracy intervals are Wilson score intervals; paired differences carry bootstrap percentile intervals~\citep{efron1979} that resample loci and, separately, clusters of loci sharing a reference gene, which is the quoted interval. Comparisons against each comparator on the primary split are reported unadjusted and labelled; the primary hypothesis is an improvement over the evidence ordering the model receives. All rates were computed from counts and rounded once. The dependence between traits and genomic regions beyond shared reference genes is not modelled.

\subsection{Data and code availability}\label{sec:data-and-code-availability}

The V2G development repository is \url{https://github.com/bowang-lab/V2G_Agentic} and is currently private. The prepared reproduction package contains the frozen benchmark and replicate-cohort extension, cached predictions from both backbones, all nine prompt-ablation runs, consensus genes, comparator calls, candidate-gene TSS coordinates, phenotype category maps, evaluation code and pinned package versions. Its fifteen-step offline workflow reconstructs the 43 cited outputs without additional model calls and checks numerical agreement for tables and pixel agreement for figures. The workflow requires the pinned Python environment and writes regenerated outputs separately from the stored results; the stored outputs can also be checked without regeneration. The authors checked freshly unpacked archives on aarch64 and x86-64; these were not independent reproductions. Live Open Targets queries are optional and disabled by default because the API can change.

The full reference datasets are available from their original providers. The release materials record the source URLs and versions and include the derived inputs and outputs needed for the reported analyses. The reference resources comprise MGI, HPO, MP and HP ontologies, GENCODE v38 TSS coordinates, 1000 Genomes LD panels, GTEx v8 eQTLs and gnomAD frequencies.

The accompanying demonstration container (v1.1; approximately 2.0~GB per architecture) includes the pipeline and interface source, reference-data slices for two loci, four example GWAS inputs, documentation, pinned package lists and verification procedures. It is prepared for linux/amd64 and linux/arm64, tested on both architectures, and distributed under the MIT licence with the original terms retained for third-party data. The demonstration runs on the bundled examples without downloading the full reference set; a full deployment requires approximately 100~GB of reference data. An archival release is planned under \href{https://doi.org/10.5281/zenodo.23489183}{\nolinkurl{doi:10.5281/zenodo.23489183}}; the public deposit is not yet available. These materials follow published guidance on code availability~\citep{nature2018code}.

\FloatBarrier

\clearpage
\onecolumn
\newgeometry{left=0.65in,right=0.65in,top=0.75in,bottom=0.75in,headheight=0pt,headsep=0pt}
\pagestyle{plain}
\appendix
\setcounter{table}{0}
\renewcommand{\tablename}{Supplementary Table}
\renewcommand{\theHtable}{S\arabic{table}}
\section*{Supplementary Information}\label{sec:supplementary}

\noindent\emph{\PaperTitle}

\emph{Supplementary analyses use the frozen benchmark and cached agent runs.}

\subsection*{Supplementary Notes}

\subsubsection*{{Supplementary Note 1 | Construction of the benchmark candidate sets}}\label{note:1}

The released benchmark applies three construction rules to the candidate sets, and every result in this study is scored on it.

Force-added genes. A reference gene that the evidence pipeline did not retrieve is added to the candidate set so that the locus remains scorable and carries an explicit force\_\allowbreak{}added flag and no composite score. A placeholder score would label such genes by their own missingness: a supervised model given only the pattern of missing features reaches Precision@1 of 1.000 on exactly those loci. Force-added and naturally retrieved loci are therefore reported separately throughout rather than pooled. Force-added loci are identified by the exact force-add signature, which occurs at 114 held-out loci; counting loci whose reference gene lacks either distance or eQTL evidence gives 237 and overstates them.

Open Targets L2G. Genuine locus-to-gene predictions were fetched from the Open Targets Platform API (Note 2) and stored in their own column, l2g\_\allowbreak{}ot, with l2g\_\allowbreak{}ot\_\allowbreak{}source recording whether the credible set came from the locus's own study; they were never folded into the composite score. Open Targets L2G scores 0.493 on the 219 loci where it predicts, 0.365 across all 296 when its abstentions are counted as failures, and 0.763 on the primary split. Crediting it with the first stored candidate at the 77 uncovered loci gives 0.422 across the 296, which describes a combined method, not Open Targets, and is labelled as such wherever it appears.

Provenance of the composite scores. The v2g\_\allowbreak{}origin field records the quantity of each composite score: 2,125 copied from an Open Targets L2G value, 2,083 derived from distance, and 309 force-added rows with no evidence-derived score.

\subsubsection*{{Supplementary Note 2 | Open Targets locus-to-gene retrieval and matching}}\label{note:2}

Locus-to-gene scores were fetched from the Open Targets Platform GraphQL API (v4). For each lead variant the API was queried for credible sets, and the L2G predictions attached to each credible set were read. If a credible set derived from the task's own study was available, it was used, which was the case for 72 of the 296 held-out loci; otherwise, the maximum score per gene across the other GWAS credible sets reported for that variant was taken. Candidates with no prediction are ranked last.

Coverage is itself a result and is reported as such. Open Targets returns at least one prediction at 219 of the 296 held-out loci and scores every candidate at 102. Its precision @1 is 0.493, where it predicts, and 0.735 where it scores the full candidate set; counting its 77 abstentions as failures gives 0.365 across all 296. Both estimands are reported throughout, because quoting either alone misrepresents a method that is permitted to decline.

The cross-study fallback (147 loci) takes the best score for a gene across credible sets of other studies and traits at the same variant, which is a different comparison from query-trait-specific L2G; the accuracy is 0.500 on same-study and 0.490 on cross-study loci. The query requested the first 50 credible sets per variant without further pages. The stored outputs did not distinguish a failed request from an empty result, and the API version did not identify the data or model release. Therefore, a missing record indicates a lack of returned prediction, not calibrated abstention. The third quantity, Open Targets with the first stored candidate as a fallback where it returns nothing (0.422 across 296; the fallback is correct at 17 loci), is reported only labelled as a combined method.

The l2g\_\allowbreak{}ot\_\allowbreak{}source field is set per locus rather than per candidate; therefore, it is populated even for candidates whose l2g\_\allowbreak{}ot value is null.

\subsubsection*{{Supplementary Note 3 | Why comparators are reported separately on naturally retrieved loci}}\label{note:3}

Every comparison in this study separates the 182 loci whose gold gene was retrieved by the evidence pipeline from the 114 loci where the reference gene had to be added to the candidate set. In the latter, the gold gene is placed at the bottom of the candidate set by construction, so any method that reads the candidate ordering is penalized there for a property of the benchmark rather than the method.

Supplementary Table 3 shows how much this matters. Across the loci scored by each comparator, no paired accuracy difference from MAGMA, TWAS, or cS2G was detected (P = 0.23, 0.15, and 0.18). Restricted to the naturally retrieved loci at which each answers, the pipeline exceeds all three: MAGMA 0.818 against 0.576 (19 wins, 3 losses, P = 8.6 \ensuremath{\times} 10\textsuperscript{\ensuremath{-}}\textsuperscript{4}), TWAS 0.808 against 0.538 (P = 1.3 \ensuremath{\times} 10\textsuperscript{\ensuremath{-}}\textsuperscript{3}), and cS2G 0.763 against 0.633 (P = 2.9 \ensuremath{\times} 10\textsuperscript{\ensuremath{-}}\textsuperscript{3}).

The reason the pooled comparison hides this is worth stating plainly, because it runs against our own interest. On force-added loci, the ordering reverses: MAGMA reaches 0.244 there and the pipeline only 0.067. Gene-level association methods score genes from variant statistics within gene boundaries and do not consult the candidate ordering at all; therefore, a gold gene buried at the bottom of that ordering costs them far less than it costs any method that reads it. Pooling the two sets of loci therefore averages a genuine advantage against an artifact and reports neither.

\subsubsection*{{Supplementary Note 4 | Mapping traits and phenotypes to organ-system categories}}\label{note:4}

Trait names and phenotype terms were mapped to a shared vocabulary of 18 organ-system categories. Trait categories were assigned by keyword rules applied to the trait list, with the rule that fired was recorded for each trait. Phenotype categories were assigned by rolling every annotated ontology term up to the top-level system terms of its ontology (the direct children of MP:0000001 for mouse phenotypes and of HP:0000118 for human phenotypes, read from the ontology files themselves) and mapping those to the same vocabulary. Categories carrying no organ-system information (cellular, embryonic, prenatal, constitutional, and "normal phenotype") were dropped explicitly rather than silently.

Both maps were constructed from the trait list and ontology files alone, without reference to predictions or reference genes, and are released as dated CSV files (2026-10-01). The file dates alone cannot show that the rules were fixed before any prediction was examined.

MGI annotations cover spontaneous, induced, and engineered alleles; no allele-class filter is applied, so these are mouse mutant phenotypes rather than knockout phenotypes specifically.

Mouse mutants are strongly pleiotropic: across annotated genes the median is five of the 18 categories and the mean 6.8, and across the benchmark's candidate genes specifically the median is 7 and the mean 7.8. Absolute support rates are therefore uninformative, and only within-locus differences between arms are interpreted.

\subsubsection*{{Supplementary Note 5 | The demonstration container image: contents, limits, data boundary and verification}}\label{note:5}

\emph{\textbf{The application is distributed as a container image}}. The demonstration build (v2g-app-demo, tag v1.1; linux/amd64 and linux/arm64; about 2.0 GB each; Python 3.11.17 with Streamlit 1.65.0 pinned, and the complete package list frozen from the built image and included in the record; code under the MIT licence) bundles slices of the reference data so that the entire pipeline runs end to end on two loci without any download. It carries the application's pipeline library; the benchmark runs evaluated in this study were made by the benchmark runner with a narrower prompt (Methods), and their outputs are in the reproduction archive.

\textbf{Contents}. Phased 1000 Genomes genotypes for all five superpopulations (GRCh38; AFR 660, AMR 347, EAS 504, EUR 503 and SAS 489 samples) for the APOB window on chromosome 2 and the 8q24 window on chromosome 8; GTEx v8 significant variant-gene pairs in both windows; the complete GENCODE v38 transcription-start-site table; the paired GRCh37/GRCh38 site list used for build inference and liftover; and the GWAS Catalog trait and EFO database. Four example inputs are included: two cohorts per trait, each restricted to one locus and each with a provenance file: LDL cholesterol from GLGC 2021 (GCST90239658, GRCh38) and GLGC 2013 (GCST002222), and prostate cancer from UK Biobank 2024 (GCST90428117, supplied in GRCh37 so that build inference and liftover are exercised) and PRACTICAL 2018 (GCST006085). The two LDL cohorts are not independent: the 2013 study is part of the 2021 meta-analysis. The two prostate cohorts come from different consortia; their independence has not been established.

\textbf{The nine steps}. Upload (parse, infer build, offer conversion); linkage disequilibrium; fine-mapping; annotation, with per-population allele frequencies; QTL mapping and colocalization; evidence combination, ranking, and optional Mendelian randomization; drug-target scoring; reasoning; report. Each step has its own run control, and most can be skipped, with each skip stating what is lost. A study is saved with what actually ran rather than with the interface's defaults: the build that was detected and how many variants were converted, the LD panel, the fine-mapping method at each locus and the sample size it used, which model at which host answered, and whether the ranking came from the model or from memory.

\textbf{Limits}. Uploads are disabled, only GWAS summary statistics are accepted, only the bundled GTEx slice is queried for expression QTLs, and LD-matrix upload is disabled. Allele frequencies come from the bundled 1000 Genomes panels; gnomAD frequencies require local gnomAD tables for full deployment. Each limit is switched off explicitly rather than left to fail, and each is lifted by one environment variable once the full reference set is present. That set, roughly 100 GB of 1000 Genomes panels, GTEx v8, gnomAD and GENCODE, is published by others under their own terms and is not redistributed; a script in the image lists every component, its size, and its source.

\textbf{What leaves the deployment}. The summary-statistics file is read inside the container and never leaves it. Steps 4 to 7 query public services with variant coordinates and alleles or with gene symbols: Ensembl VEP, the Open Targets Platform, the gnomAD and GTEx portals, STRING and DisGeNET. The optional GWAS Catalog browser sends trait and study identifiers to the EBI GWAS Catalog. Step 8 sends the candidate genes, the locus record (lead variant, the top posterior inclusion probability, and the variant that carries it, whether the lead is in a credible set, the other candidate and non-coding genes at the locus), and the assembled per-gene evidence to the model endpoint the user configures; with memory enabled, it also sends recalled anchor templates, each labelled same-trait or cross-trait. In a full deployment, a PDF input read with model-assisted extraction also sends the document's text to that endpoint (PDF input is disabled in the demonstration image). Every model request is logged before it is sent as host, model, message count, payload size, and whether the destination was local or external, with no key and no prompt text. Steps 1 to 3 run with networking disabled, and later steps then report the sources they could not reach rather than omit them silently.

\textbf{Memory}. The image ships with the EviMem and CAVE memory components enabled, the application's default; memory can be set to none in the interface. After a model-reasoned run in which every lookup succeeded, EviMem stores one anchor per locus: the top-ranked gene with a 16-dimensional evidence fingerprint, of which the application fills 12 (composite score, druggability, colocalization posterior, Mendelian randomization p-value, estimate and significance, direction of effect, the colocalization, variant-consequence and variant-to-gene flags, lead GWAS p-value and number of expression-QTL tissues). A later study recalls anchors by cosine similarity of fingerprints and labels each as same-trait or cross-trait. The model is skipped and the ranking transferred only when the best similarity reaches a threshold that rises as the fingerprint gets sparser (0.93 with eight or more informative dimensions, up to 0.99 with two, the minimum) and the anchor's gene is a candidate at the new locus; otherwise, the recalled anchors are passed to the model as labelled context. No transfer is made when a variant-consequence lookup failed, because the fingerprint cannot distinguish a failed lookup from an absent consequence, and a transferred run writes no new anchor. CAVE records the evidence, claims, contracts, and adjudication behind each call. Two limits apply: the fingerprint's consequence field is locus-level, so two studies whose leads differ only in coding consequence can match; and the benchmark's locus records carry fewer fields, so a similarity in the application is not on the benchmark's scale. This component is not part of the evaluation reported in this paper, whose runs make one model call per locus with no memory.

\textbf{Verification}. A headless script drives both demonstration studies of each trait through parse, build inference, liftover, LD, fine-mapping, annotation, ancestry frequencies and scoring, then the memory and CAVE evidence, claim, contract and adjudication path, and asserts on what was recorded: 167 checks, including that all 377 GRCh37 variants of the UK Biobank file convert, the number of fine-mapping signals and the sum of posterior probabilities at each locus, that the saved method label and the report agree with the per-locus record, that an unconverted GRCh37 study is flagged and its report says so (the interface refuses to compute LD for it), that each lead variant has frequencies in all five superpopulations, that the memory transfer decision follows the similarity gate and does not depend on the order recalled anchors arrive in, and that every transferred claim carries a rationale. A second script checks the host architecture, the imports, the bundled data, and the service health (36 checks). Both were run inside the image on an arm64 machine and an x86-64 machine by the authors, from the documented commands; this is an installation and run check on a second architecture, not an independent reproduction. The library behind the application is covered by a regression test suite (675 tests) that gates every image build. The deposited documentation lists the limitations of this build.

\subsubsection*{{Supplementary Note 6 | Evaluation choices behind the reported numbers}}\label{note:6}

\textbf{Predictions}. One gene per locus is selected by a plurality vote over the five seeds with label-free tie-breaks, and the selected genes are released. A five-seed mean of correctness thresholded at 0.5 is not used; it requires the reference labels and names of no gene (on this benchmark, it gives the same per-locus correctness).

\textbf{Functional analysis}. Phenotype support is computed on the consensus GPT-5.2 genes, the paper's primary predictions (144 of 296 correct), using the same phenotype-support analysis throughout. The primary background is label-free; a background that excludes the reference genes depends on the labels and is reported only marked as such.

\textbf{Nearest gene}. Nearest-gene results use the GENCODE transcription-start-site distance described in the Methods section. The benchmark's stored distance field is present at only 97 of 296 test loci, and ranking by it falls back to the first stored candidate, the evidence-ordering top, at the other 199; on the primary split that definition gives 0.696 against 0.615 by TSS, and it is shown only labelled.

\textbf{Baselines and overlap}. The membership baselines are named for what they compute, evidence-free frequency controls are included, and the train/test overlap is measured by kind (Supplementary Table 8).

\textbf{What the model saw}. In the evaluated runs, each gene's prompt line carried the composite score, binary eQTL, colocalization, coding-consequence flags, and a column labelled L2G that duplicated the composite score; no distances, tissue counts, posteriors, or Mendelian randomization estimates. Real colocalization values exist in the benchmark at 17 of 296 test loci and were computed from genome-wide-significant GWAS variants and GTEx significant pairs only.

\textbf{Not done, and stated as limitations}: re-splitting by association and retraining the comparators; a comparison with general biomedical agents; and checks of the fine-mapping, colocalization, and Mendelian randomization inputs per locus. Prompt experiments that withhold the score and the ordering, give gene names and the trait alone, or attach each gene's evidence to another gene's symbol are in Note 9 (three seeds each); a non-language-model reranker on the prompt's inputs is in Note 8.

\subsubsection*{{Supplementary Note 7 | The prompt as evaluated}}\label{note:7}

The benchmark runs evaluated in the main text made one language-model call per locus. The stored runs retain the parsed response but not the prompt text; the two prompts below are reproduced from the runner code (\_\allowbreak{}SYSTEM\_\allowbreak{}PROMPT and \_\allowbreak{}build\_\allowbreak{}user\_\allowbreak{}prompt in the ablation runner), and the rendered example is the prompt that the code produces for the locus of Figure 1. The stored runs were made on an earlier build of the same candidate sets, identical in membership and ordering, in which a force-added gene carried a placeholder composite score of 0.1 (rendered as v2g=0.10) and the L2G field duplicated the composite score; therefore, the prompts contain those values, while every result in this study is scored on the released benchmark. The evaluated configuration ("no\_\allowbreak{}memory") passes no memory context; therefore, the block that other configurations append is absent.

\textbf{System prompt, verbatim:}

\emph{You are an expert biomedical geneticist specializing in GWAS causal gene identification and drug target prioritization. Given candidate genes at a GWAS locus with V2G scores, druggability grades, and evidence flags, rank them by the likelihood of being the primary causal therapeutic target. Respond ONLY with valid JSON (no markdown, no extra fields) in this exact format: \{"ranked\_\allowbreak{}genes": [\{"gene": "SYMBOL", "rank": 1, "causal\_\allowbreak{}probability": 0.0, "reasoning": "one sentence only"\}], "biological\_\allowbreak{}mechanism": "one sentence"\} Rules: causal\_\allowbreak{}probability in [0-1]; reasoning MUST be one sentence per gene (\ensuremath{\leq}20 words); biological\_\allowbreak{}mechanism one sentence; NO key\_\allowbreak{}evidence, NO confidence fields. List ALL candidates in rank order.}

\textbf{User prompt template}. Candidates appear in the order of the benchmark record, which is the descending composite score. Each gene is one line: "\textbullet{} GENE: CIR=<cir>, druggability=<tier>, evidence=[<fields>]", where <fields> lists, in this order and only where present, v2g=<composite score to two decimals> (omitted when zero), eQTL, coloc, coding\_\allowbreak{}variant, MR\_\allowbreak{}\ensuremath{\beta}=<estimate> and L2G=<value>. The CIR and druggability fields were unpopulated in the benchmark and read 0.00 and 1 for every gene; the MR field never appears because the benchmark holds no Mendelian randomization estimate; L2G appears for the 2,125 of 4,517 candidates whose earlier build held a value, which duplicated the composite score. The prompt begins "Trait: <trait>" and "Candidate genes (pre-scored by V2G pipeline):" and ends "Rank all genes by causal probability. Return compact JSON only."

\textbf{Rendered for 2:21155063 A>G (LDL cholesterol; locus of Figure 1 and 4b), exactly as sent}:

\textbf{Trait}: low-density lipoprotein cholesterol measurement\\
\textbf{Candidate genes (pre-scored by V2G pipeline):}

\begin{itemize}
\item TDRD15: CIR=0.00, druggability=1, evidence=[v2g=0.94, eQTL, coloc]
\item APOB: CIR=0.00, druggability=1, evidence=[v2g=0.78, eQTL, coloc]
\item LDAH: CIR=0.00, druggability=1, evidence=[v2g=0.34, eQTL, coloc]
\item GDF7: CIR=0.00, druggability=1, evidence=[v2g=0.02, eQTL, coloc]\\
Rank all genes by causal probability. Return compact JSON only.
\end{itemize}

At this locus, the four candidates differ only in the composite score; the eQTL and colocalization flags are set for all four, no coding consequence is flagged, and no L2G value is present. Nothing in the prompt names a tissue; therefore, the "lipid eQTL" in the stored rationale for APOB (Figure 4b) is an inference from the trait name, not a retrieved measurement. Calls are capped at 800 output tokens.

\subsubsection*{{Supplementary Note 8 | A classifier restricted to the prompt's inputs}}\label{note:8}

To test whether the gain over the evidence ordering could be a learnable function of the fields the model was shown, two classifiers were trained on exactly those fields (the composite score and the eQTL, colocalization, and coding flags) using the naturally retrieved training loci, the same protocol as the gradient-boosted comparator in Table 1: a gradient-boosted model (300 trees, depth 3, learning rate 0.05) and a standardized logistic regression. Each scores candidates per locus and the top-scored gene is the call. Supplementary Table 10 gives Precision@1 by split with Wilson intervals, paired against the GPT-5.2 consensus by exact McNemar test. Both classifiers fall below the evidence ordering itself on every split, so the retrieved flags are not a sufficient explanation of the model's gain; Note 9 separates anchoring from recall. These comparisons use the frozen benchmark and cached predictions; they require no additional model calls.

\subsubsection*{{Supplementary Note 9 | Prompt ablations: names only, flags without the score, and evidence permuted between genes}}\label{note:9}

Three further GPT-5.2 variants were run over the 296 held-out loci, three seeds each (2,664 calls), with the benchmark runner, the same client settings, benchmark tasks, and candidate lists as the primary runs, differing only in the prompt. Names only: the trait and the gene symbols in a shuffled order, nothing else. Flags only, shuffled: each symbol with its eQTL, colocalization, and coding-consequence flags, no composite score, no L2G value, and neither of the constant CIR and druggability fields, again shuffled. Evidence permuted: the evaluated prompt unchanged in text and in order, with the gene symbols reassigned among the locus's evidence rows by a derangement, so that no gene keeps its own evidence and the gene shown first, with the highest score, is never the gene whose evidence placed it first; because the scores, the order, and the framing of the candidates as pre-scored are unchanged, this is a stress test of anchoring under contradictory inputs, not an estimate of the contribution of the evidence content. The shuffle and derangement are drawn from the seed and the locus identifier; the shown order, the permutation, the prompt as sent, and the API usage record are stored with each run, and the stored responses are scored exactly as the primary runs (Supplementary Table 11). No run recorded an error; 2 of the responses named a symbol outside the candidate set (1 in flags only, shuffled (seed 0); 1 in names only, shuffled (seed 2)), each scored as incorrect, as in the primary runs. The client sends GPT-5.2 neither a temperature nor a reasoning-effort parameter; the usage records of the 2,072 calls that carry one report 0 reasoning tokens and finish reason stop at every call, under a completion cap of 800 tokens.

System prompt of the names-only run, verbatim (the flags-only system prompt differs only in saying "listed in no particular order, each with its evidence flags"; the permuted run uses the evaluated system prompt of Note 7):

You are an expert biomedical geneticist specializing in GWAS causal gene identification and drug target prioritization. Given the candidate genes at a GWAS locus, listed in no particular order and with no further information, rank them by likelihood of being the primary causal therapeutic target. Respond ONLY with valid JSON (no markdown, no extra fields) in this exact format: \{"ranked\_\allowbreak{}genes": [\{"gene": "SYMBOL", "rank": 1, "causal\_\allowbreak{}probability": 0.0, "reasoning": "one sentence only"\}], "biological\_\allowbreak{}mechanism": "one sentence"\} Rules: causal\_\allowbreak{}probability in [0-1]; reasoning MUST be one sentence per gene (\ensuremath{\leq}20 words); biological\_\allowbreak{}mechanism one sentence; NO key\_\allowbreak{}evidence, NO confidence fields. List ALL candidates in rank order.

User prompts as sent at seed 0 for 2:21155063 A>G (LDL cholesterol; the locus of Figure 1): names only, flags only, and evidence permuted. The stored order of this locus is TDRD15, APOB, LDAH, GDF7; the shuffled order shown to the first two was APOB, GDF7, LDAH, TDRD15. In the permuted prompt, the rows keep the stored order and scores, and the symbols move: APOB shown with the evidence of TDRD15, LDAH shown with the evidence of APOB, GDF7 shown with the evidence of LDAH, and TDRD15 shown with the evidence of GDF7. All three runs ranked APOB, LDAH, GDF7, and TDRD15; the permuted run's top gene is the first-shown row, which carries the score of TDRD15.

Trait: low density lipoprotein cholesterol measurement

Candidate genes (in no particular order):

\begin{itemize}
\item APOB
\item GDF7
\item LDAH
\item TDRD15
\end{itemize}

Rank all genes by causal probability. Return only compact JSON.

Trait: low density lipoprotein cholesterol measurement

Candidate genes (in no particular order):

\begin{itemize}
\item APOB: evidence=[eQTL, coloc]
\item GDF7: evidence=[eQTL, coloc]
\item LDAH: evidence=[eQTL, coloc]
\item TDRD15: evidence=[eQTL, coloc]
\end{itemize}

Rank all genes by causal probability. Return compact JSON only.

Trait: low density lipoprotein cholesterol measurement

Candidate genes (pre-scored by the V2G pipeline):

\begin{itemize}
\item APOB: CIR=0.00, druggability=1, evidence=[v2g=0.94, eQTL, coloc]
\item LDAH: CIR=0.00, druggability=1, evidence=[v2g=0.78, eQTL, coloc]
\item GDF7: CIR=0.00, druggability=1, evidence=[v2g=0.34, eQTL, coloc]
\item TDRD15: CIR=0.00, druggability=1, evidence=[v2g=0.02, eQTL, coloc]
\end{itemize}

Rank all genes by causal probability. Return compact JSON only.

\textbf{Results (mean of 3 seeds, range in parentheses)}. On the primary split, the names-only prompt scores 0.864 (0.859-0.867) and the flags-only prompt 0.862 (0.852-0.867), against 0.795 (0.778-0.807) for the full prompt, 0.793 (107/135) for the five-seed consensus, and 0.711 (96/135) for the evidence ordering; the permuted prompt scores 0.284 (0.267-0.311). On the naturally retrieved loci: names only 0.839 (0.835-0.846), flags only 0.832 (0.813-0.841), full prompt 0.758 (0.742-0.769), consensus 0.753 (137/182), permuted 0.337 (0.319-0.357). On the force-added loci, where the reference gene was never retrieved and entered the candidate set with a placeholder score, the names-only prompt selects it at 0.594 (0.588-0.605) and the flags-only prompt at 0.456 (0.430-0.482), against 0.076 (0.053-0.105) for the full prompt and 0.061 (7/114) for the consensus; the permuted prompt selects it at 0.439 (0.430-0.456), for a mechanical reason: the derangement moves the reference symbol off its placeholder row onto a scored one. Over all 296 loci the permuted prompt scores 0.376 (0.372-0.385) against 0.395 (117/296) for the evidence ordering; this total is the sum of opposite effects on natural and force-added loci, not an equivalence. On the 97 primary loci whose reference gene and trait never occur together among the training reference genes (the overlap-free subset): names only 0.832 (0.825-0.835), flags only 0.828 (0.814-0.835), full prompt 0.794 (0.783-0.804), consensus 0.794 (77/97), evidence ordering 0.701 (68/97), permuted 0.272 (0.247-0.299). Names only against flags only, same seed: on the primary split 2-3 loci to 2, P = 1.000; on the force-added loci 12-18 loci to 0, P <= 4.9e-04. At the loci where the full prompt corrects the evidence ordering, the names-only prompt of the same seed is also correct at 13 of 13, 12 of 12, 9 of 9 primary loci per seed and 27 of 29, 33 of 34, 25 of 26 loci per seed over the whole test set. One further full-prompt pass was made in October 2026 beside the ablations, as a control for drift of the served model since the June primary runs (same runner, prompt, client and benchmark tasks): Precision@1 0.756 (102/135) on the primary split and 0.470 (139/296) over all loci, below the June seeds scored here (0.778-0.807 and 0.483-0.510). Paired with each June seed on the primary split (June seed wins to control wins): 7 to 0, P = 0.016; 6 to 0, P = 0.031; 3 to 0, P = 0.250. Against this October pass the names-only prompt of the same month wins 21-22 loci to 7 (P <= 0.013). The control is reported in Supplementary Table 11 as a single repetition and enters no primary comparison.

\textbf{Diagnostics per seed}. With the full prompt, the top-ranked gene is the first gene of the evidence ordering at 249-257 of 296 loci; with names only, it coincides with that gene at 140-141, and the first gene shown is chosen at 67-84 of 296 (expected 0.273 of loci if position decided). With the evidence permuted, the top-ranked gene is the first-shown row, carrying another gene's symbol, at 231-238 of 296, and the gene whose own evidence had been first at 15-24; the reference symbol landed on the first row at 62-71 loci. Paired exact McNemar tests are computed separately for each seed, with no pooling across seeds; the full results are released with the analysis. The model therefore names these reference genes from the trait and the symbol alone; the three flags add nothing on the primary split and lower accuracy on the force-added loci; and when a scored ordering is shown it follows the top-scored row at most loci (231-238 of 296 with the evidence permuted), whichever symbol it carries, so the ordering with its score anchors it to the first candidate (Discussion).

\textbf{\\
}

\begin{landscape}
\subsection*{Supplementary Tables}
\begingroup
\fontsize{8.0}{9.5}\selectfont
\setlength{\tabcolsep}{3pt}
\renewcommand{\arraystretch}{1.1}
\setlength{\LTcapwidth}{\linewidth}

\endgroup

\end{landscape}
\end{document}